\documentclass[10pt,journal]{IEEEtran}
\usepackage{amsmath}
\usepackage{comment}
\usepackage{amssymb} 
\usepackage{graphicx} 
\usepackage{cite}
\usepackage{algorithm}
\usepackage{algorithmic}
\usepackage{array}
\usepackage{float} 
\usepackage{multirow} 
\usepackage{booktabs}
\usepackage[font=small]{caption}
\usepackage[export]{adjustbox}
\usepackage[caption=false,font=footnotesize]{subfig}
\usepackage[table,dvipsnames]{xcolor}

\newcolumntype{C}[1]{>{\centering\arraybackslash}p{#1}}

\begin{document}

\title{ Hybrid MIMO in the Upper Mid-Band:\\
Architectures, Processing, and Energy Efficiency}
\author{Marouan Mizmizi, Ahmed Alkhateeb, Umberto Spagnolini
\thanks{Marouan Mizmizi and Umberto Spagnolin are with the Department of electronics, Information and Bioengineering, Politecnico di Milano, 20133, Milano, Italy, Emails: marouan.mizmizi@polimi.it, umberto.spagnolini@polimi.it}
\thanks{Ahmed Alkhateeb is with the School of Electrical, Computer, and Energy Engineering, Arizona State University. Email: alkhateeb@asu.edu}
\thanks{This work was partially supported by the European Union - Next Generation EU under the Italian National Recovery and Resilience Plan (NRRP), Mission 4, Component 2, Investment 1.3, CUP D43C22003080001, partnership on “Telecommunications of the Future” (PE00000001 - program “RESTART”)}}

\maketitle

\begin{abstract}
As 6G networks evolve, the upper mid-band spectrum (7 GHz–24 GHz), or frequency range 3 (FR3), is emerging as a promising balance between the coverage offered by sub-6 GHz bands and the high-capacity of millimeter wave (mmWave) frequencies. This paper explores the structure of FR3 hybrid MIMO systems and proposes two architectural classes: \textit{Frequency Integrated (FI)} and \textit{Frequency Partitioned (FP)}. FI architectures enhance spectral efficiency by exploiting multiple sub-bands parallelism, while FP architectures dynamically allocate sub-band access according to specific application requirements. Additionally, two approaches—fully digital (FD) and hybrid analog-digital (HAD)—are considered, comparing shared (SRF) versus dedicated RF (DRF) chain configurations. Herein signal processing solutions are investigated, particularly for an uplink multi-user scenario with power control optimization. 

Results demonstrate that SRF and DRF architectures achieve comparable spectral efficiency; however, SRF structures consume nearly half the power of DRF in the considered setup. While FD architectures provide higher spectral efficiency, they do so at the cost of increased power consumption compared to HAD. Additionally, FI architectures show slightly greater power consumption compared to FP; however, they provide a significant benefit in spectral efficiency (over $4 \times$), emphasizing an important trade-off in FR3 engineering.

\end{abstract}


\IEEEpeerreviewmaketitle

\section{Introduction}
\begin{figure}[b]
    \centering
    \includegraphics[width=0.75\linewidth]{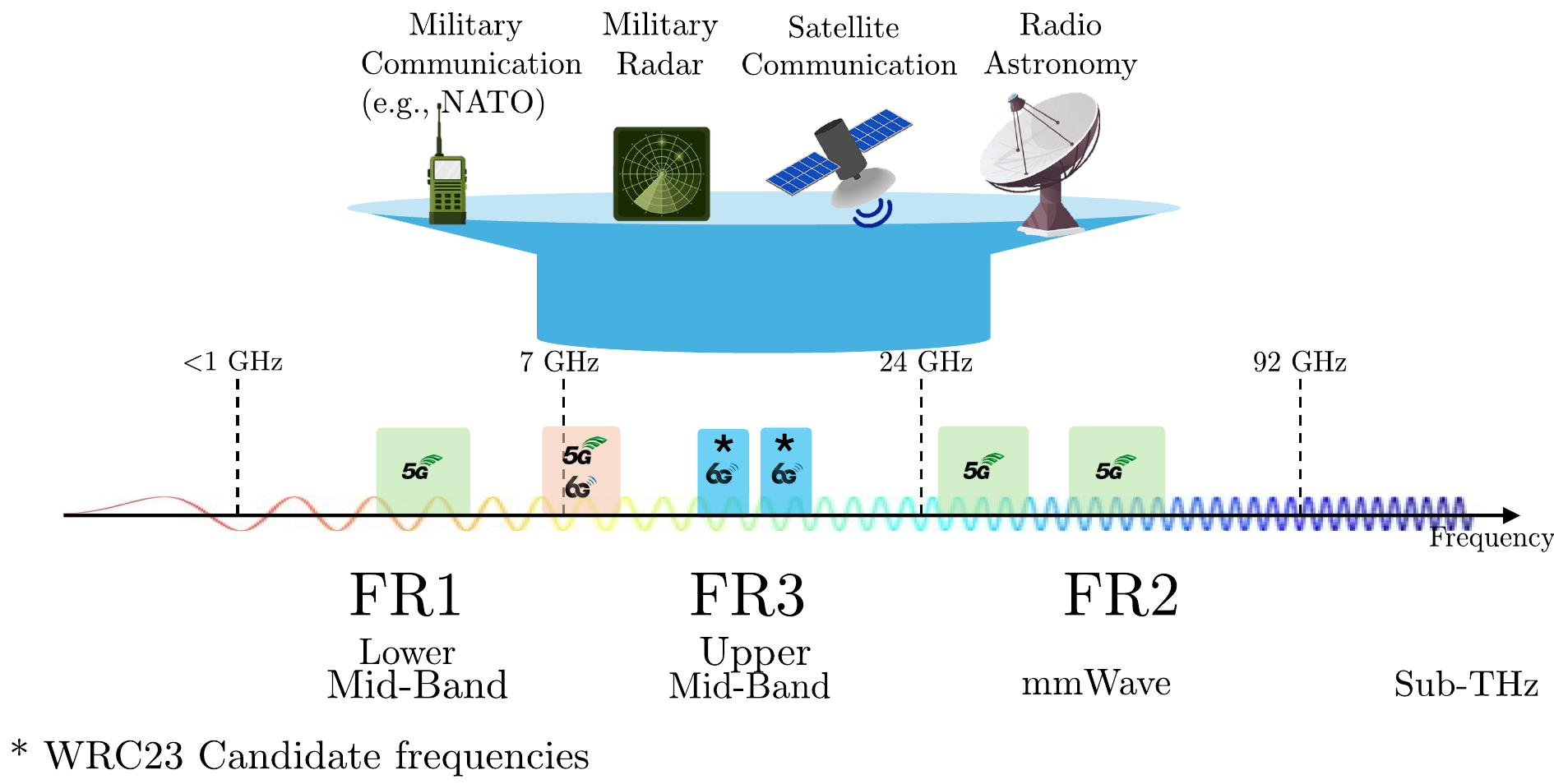}
    \caption{FR3 Spectrum and candidate frequencies for 5G/6G and 6G}
    \label{fig:spectrum}
\end{figure}

As 6G networks begin to take shape, there is much anticipation on the transformative applications enabled, ranging from ultra-reliable communications to immersive augmented reality and high-speed vehicle-to-vehicle interaction \cite{giordani2020toward}. One of the key challenges for 6G is to provide a solution that balances the need for high capacity with coverage in diverse environments \cite{10634051}. This challenge has led to growing interest in frequency range 3 (FR3), a new spectrum proposed during the $23^\mathrm{th}$ World Radio Conference (WRC) \cite{rspg_report_wrc23}, which spans from 7 GHz to 24 GHz, as depicted in Fig. \ref{fig:spectrum}.

Millimeter wave frequencies (mmWave) were initially considered a promising candidate for mobile broadband due to their large available bandwidth. However, their poor propagation characteristics, including high indoor-to-outdoor attenuation and severe path loss, have significantly limited their applicability in mobile communications \cite{mmwave_underutilized}. As a result, FR3 also referred to as the upper mid-band, has gained attention as a compromise between highly congested sub-6 GHz bands (FR1) and limited coverage of mmWave frequencies (FR2) \cite{shakya2024comprehensive}. The drawback of FR3 is the fragmentation of the spectrum into multiple subbands \cite{raviv2024multi, zhang2024new}. 

Despite its advantages, several challenges must be addressed to fully exploit FR3’s potential. One significant issue is spectrum sharing with existing services such as satellite communications, military radar, and passive radio astronomy. For instance, the 10.7–12.75 GHz band is heavily utilized for satellite TV, while the 14–18 GHz range is allocated for high-resolution military radar \cite{ECC}. To ensure coexistence among these services, advanced mechanisms such as spectrum sensing and interference suppression are necessary \cite{kang2024cellular}.

On the hardware side, the wide and interleaved bandwidths in FR3 and the varying carrier frequencies introduce considerable complexity. The design of compact and energy-efficient systems that can support these specifications pose a significant challenge \cite{10198042}. Compact multi-band antenna arrays and broadband components are essential for managing the dynamic spectrum environment. These technologies not only facilitate the simultaneous processing of multiple carriers but also support directional transmission techniques, which are crucial for minimizing interference and accommodating spectrum sharing.

A simple approach to address the challenges of FR3 involves implementing separated RF front ends with distinct RF chains for each carrier. Each chain processes RF signals independently from others, and the outputs are aggregated at the baseband processing (BB) stage. Although this architecture is the most straightforward solution, it is RF resource-intensive and inefficient due to the significant redundancy in hardware components. Nevertheless, only a few studies have been conducted in this direction. For example, the authors of \cite{10694319} developed a software-defined radio platform with frequency hopping for FR3. This architecture employs distinct antennas for each carrier, with common RF chains and a shared BB stage to process transmitted or received signals, with an RF switch used to select the active antenna and corresponding carrier. While this approach is cost-effective, it processes each carrier sequentially, rather than simultaneously, which limits the spectral efficiency that could be achieved through the simultaneous processing of multiple carriers.
A more sophisticated alternative for FR3, proposed in \cite{10198042,banerjee2024flexible}, proposes unified multiband RF front ends using wideband components to handle multiple carriers within shared RF chains. While this reduces hardware redundancy and increase spectral efficiency, it demands meticulous design and high-performance components.

\subsection*{\textbf{Contributions}}

The architectural choices in FR3 systems have a significant impact on system performance, requiring tailored signal processing strategies that align with the hardware design. However, the existing literature lacks a comprehensive exploration of these trade-offs. This paper addresses this critical gap by providing a novel classification of architectural design strategies and analyzing their performance in uplink multi-user, multi-band MIMO scenarios.

Specifically, the paper introduces two classes of multi-band access strategies: \textit{Frequency Integrated (FI)} and \textit{Frequency Partitioned (FP)}. In FI architectures, the system simultaneously accesses multiple sub-bands to optimize overall spectral efficiency. In contrast, FP architectures dynamically allocate access to sub-bands, scheduling them based on specific application requirements. Within these categories, the paper further explores various design strategies, distinguishing between fully digital (FD) and hybrid analog-digital (HAD) approaches, and also considers whether RF chains are shared (SRF) or dedicated (DRF) among sub-bands.

The main contributions of this work are as follows:
\begin{itemize}
    \item We propose a comprehensive classification of potential architectures for FR3 systems, distinguishing between FI and FP approaches. We further differentiate FD and HAD designs within each class, considering whether RF chains are shared (SRF) or dedicated (DRF) among sub-bands.
    \item For each architecture, we derive the optimal power control and beamforming strategies, specifically tailored to maximize the sum rate across all sub-bands simultaneously. This includes developing optimal beamforing strategies for hybrid MIMO systems under hybrid beamforming constraints imposed by phase-shifting networks. 
    \item  For each architecture, we comprehensively analyze the impact of key RF components, including noise figure, power consumption, and linear distortions, on the system's performance. The study specifically evaluates the effects of these impairments, providing insights into their influence on architectural choices.
\end{itemize}

The remainder of this paper is organized as follows. Sec. 2 defines the system model for the multi-user, multi-band FR3 scenario. Sec. 3 analyzes the hardware impairments, focusing on the key RF components. Sec.4 outlines the proposed architectural designs along with their classification. Se. 5 presents the optimization framework, detailing power allocation and beamforming strategies. Sec. 6 discusses the performance evaluation. Finally, Section 7 concludes the paper.

\section{System Model}

\begin{figure}[b!]
\vspace{-.5cm}
    \centering
    \includegraphics[width=0.8\linewidth]{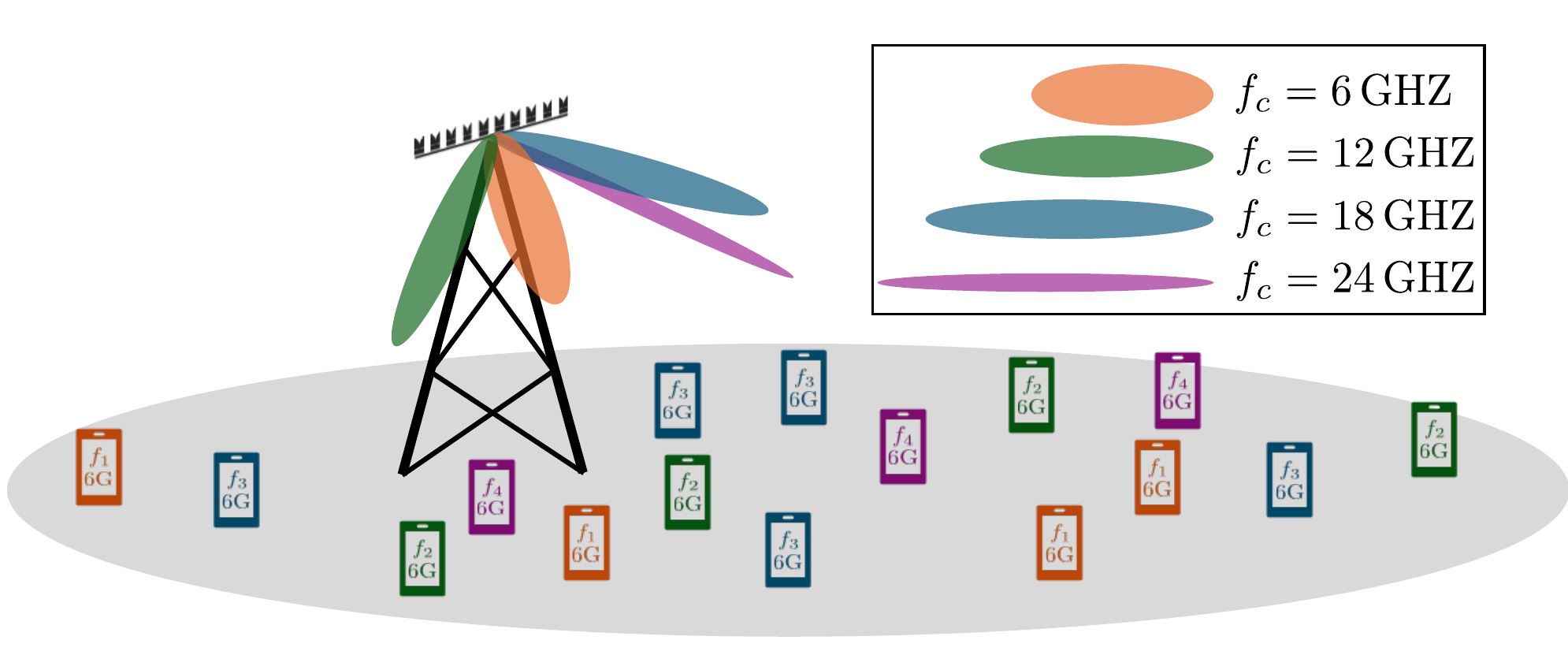}
    \caption{Multi-user multi-band reference scenario.}
    \label{fig:scenario}
\end{figure}

We consider the MIMO base station (BS), depicted in Fig. \ref{fig:scenario}, with $N_a$ antennas, operating in the FR3 and supporting multiple sub-bands with different bandwidths ($B_c$), specifically \( B_1 \) at 6 GHz, \( B_2 \) at 12 GHz, \( B_3 \) at 18 GHz, and \( B_4 \) at 24 GHz. The BS serves multiple single-antenna UEs, denoted by the set \( \mathcal{K} \), with cardinality \( |\mathcal{K}| = K \). Each sub-band \( c \), \( c \in \{1, 2, 3, 4\} \), supports a disjoint subset of UEs:
\begin{equation*}
    \mathcal{K} = \bigcup_{c=1}^{4} \mathcal{K}_c, \quad \mathcal{K}_i \cap \mathcal{K}_j = \varnothing, \; \forall i \neq j.
\end{equation*}
Considering the uplink operation (the downlink can be derived with minimal effort), let $s_k \in \mathbb{C}$ be the data symbol transmitted by the $k$th UE, such that $\mathbb{E}\left[s_k^\mathrm{*} s_\ell \right] = \lambda_k \delta[k-\ell]$, where $\lambda_k$ denotes the power transmitted by $k$th UE. 
At the BS, the signal received on sub-band \( c \) from all \( \mathcal{K}_c \) UEs is given by  
\begin{equation} \label{eq:rx_signal}
    \mathbf{y}[c] = \sum_{k \in \mathcal{K}_c} \mathbf{h}_k[c] s_k + \mathbf{n}[c],
\end{equation}
where \( \mathbf{h}_k[c] \in \mathbb{C}^{N_A \times 1} \) denotes the block fading channel vector between the BS and the \( k \)-th UE for sub-band \( c \), while \( \mathbf{n}[c] \sim \mathcal{CN}\left(\mathbf{0}, (\sigma_n^2 + \sigma_\mathrm{hwi}^2)\mathbf{I}_{N_a}\right) \) represents the additive noise, which consists of both thermal noise $\sigma_n^2$ and distortion noise $\sigma_\mathrm{hwi}^2$ due to hardware impairments. The contribution of thermal noise is
\begin{equation}\label{eq:noisepower}
    \sigma_n^2 = \Gamma  B \, 10^\frac{\mathrm{NF}}{10},
\end{equation}
where \( \Gamma = K_b T \) depending on the Boltzmann constant \( K_b \), and the reference operating temperature \( T = 290K \). \( B \) denotes the system bandwidth, while \( \mathrm{NF} \) represents the noise figure of the considered architecture.
\(\sigma_\mathrm{hwi}^2\) models the linear distortions induced by hardware impairments \cite{schenk2008rf,9226127}:
\begin{equation}
    \sigma_\mathrm{hwi}^2 = \varsigma \left\|\sum_{k \in \mathcal{K}_c} \mathbf{h}_k[c] s_k\right\|^2,
\end{equation}
where \(\varsigma\) is a parameter dependent on the quality and imperfections of the hardware components. Unlike thermal noise, \(\sigma_\mathrm{hwi}^2\) is a signal-dependent noise term that increases with the received power level, ultimately leading to performance saturation as the input signal power grows.

The signal received on the $c$th sub-band is processed at BS using the spatial combiner \( \mathbf{W}[c] \in \mathbb{C}^{N_a \times K_c}\). The design of the specific beamforming matrix depends on the hardware architecture considered, as detailed in Section \ref{sec:architecture}. For mathematical convenience, let us assume that the combination matrix changes randomly with channels $\mathbf{h}_k[c]$ such that $\mathbb{E}[\mathbf{W}^\mathrm{H}[c]\mathbf{W}[c]] = \mathbf{I}_{K_c} / K_c$. The decoded symbol for the $k$th UE can be expressed as
\begin{equation} \label{eq:combined_signal}
\hat{s}_k[c] = \mathbf{w}_k[c]^H \mathbf{h}_k[c] s_k + \underbrace{\sum_{\substack{\ell  \in \mathcal{K}_c \\ \ell \neq k}} \mathbf{w}_k[c]^H \mathbf{h}_\ell[c] s_\ell}_{\mathrm{Interference}} + \mathbf{w}_k[c]^H \mathbf{n}[c],
\end{equation}
where \( \mathbf{w}_k[c] \) is the \( k \)-th column of \( \mathbf{W}[c] \). In equation \eqref{eq:combined_signal}, the received symbol is influenced by impairments that depend on the architecture used and the interference from other UEs allocated within the same sub-band.

\subsection{Channel Model}

Modeling the channel in the upper mid-band for MIMO systems is an active area of research that still presents several open questions, particularly given the distinct propagation behaviors at various frequencies \cite{shakya2024comprehensive,zhang2024new}. The radio waves in FR3 bands are expected to travel over the same paths, resulting in the same delays and angles. However, the signal will experience different gains and phases due to wavelength-dependent interactions with the scatterers. Additionally, in large MIMO systems, the distinction between near-field and far-field regions varies carrier by carrier, adding further complexity to channel modeling.

Given that our focus is not on developing a new channel model, we adopt a far-field propagation with a cluster-based model consistent with the latest state-of-the-art approaches \cite{zhang2024new}. This model assumes common cluster locations for all carriers while recognizing that each scatterer within a cluster responds with frequency-dependent variations. In detail, the channel between the BS and any arbitrary UE in the carrier $f_c$ is
\begin{equation}
    \mathbf{h}_k [c] = \sum_{q=1}^{Q} \sum_{p=1}^{P_q} \alpha_{q,p}[c] \, \mathbf{a}(\theta_{q,p}, \varphi_{q,p}),
\end{equation}
where $Q$ and $P_q$ represent the number of scattering clusters and the related number of scatterers, respectively. $\mathbf{a}(\theta_{q,p}, \varphi_{q,p}) \in \mathbb{C}^{N_A \times 1}$ denotes the BS array response vectors,  with $\theta_{q,p}$ and $\varphi_{q,p}$ being the azimuth and elevation angles of arrival, respectively. $\alpha_{q,p}[c]$ is the complex gain associated with the $p,q$th scatterer, and is such that $\mathbb{E}\left[\left|\sum_p \alpha_{q,p}[c]\right|^2\right] = \gamma_q[c]$, with $\gamma_q[c]$ being the cluster power calculated in $dB$ based on the 3GPP model in \cite{3gpp_ts38901} as
\begin{equation}
    \gamma_q[c] = -32.4 - 21 \log_{10}(d) - 20 \log_{10}(f_c),
\end{equation}
that decreases monotonically with the carrier frequency $f_c$ and with the end-to-end distance $d$.
\section{Hardware Impact and Power Dissipation}

This section explores the role of hardware in system performance, which is closely tied to architectural design decisions. As such, providing a comprehensive, albeit general, analysis is essential.

RF impact is assessed by two key factors: \textit{power consumption} and \textit{noise figure (NF)}. Active power consumption represents the energy required for a component to function and includes both static and dynamic contributions, depending on the component's purpose and load conditions.
Architectural performance is shaped by hardware impairments, comprising a static NF and a dynamic component. The total NF in a multi-stage receiver follows Friis' formula:
\begin{equation}\label{eq:friis}
    \mathrm{NF}_{\text{total}} = \mathrm{NF}_1 + \sum_{i=2}^{N} \frac{\mathrm{NF}_i - 1}{\prod_{j=1}^{i-1} G_j}
\end{equation}
where \( \mathrm{NF}_i \) and \( G_i \) denote the noise figure and gain of the \( i \)-th stage, respectively. The low-noise amplifier (LNA) predominantly defines the first-stage noise figure, \( \mathrm{NF}_1 \). Its placement at the front of the receiver chain makes it critical, as noise added at this stage cannot be suppressed by the gain of the following stages.

Notice that this paper does not include hardware-induced nonlinearities. However, these nonlinearities are inherent in all receivers and require a careful integrated circuit design to decrease their impacts \cite{rebeiz2013wideband}. This work targets a comparative analysis of architectures, so it assumes that these effects uniformly influence all designs considered, with in-depth investigation deferred to future research.

\begin{figure}[b!]
    \centering
    \includegraphics[width=0.7\linewidth]{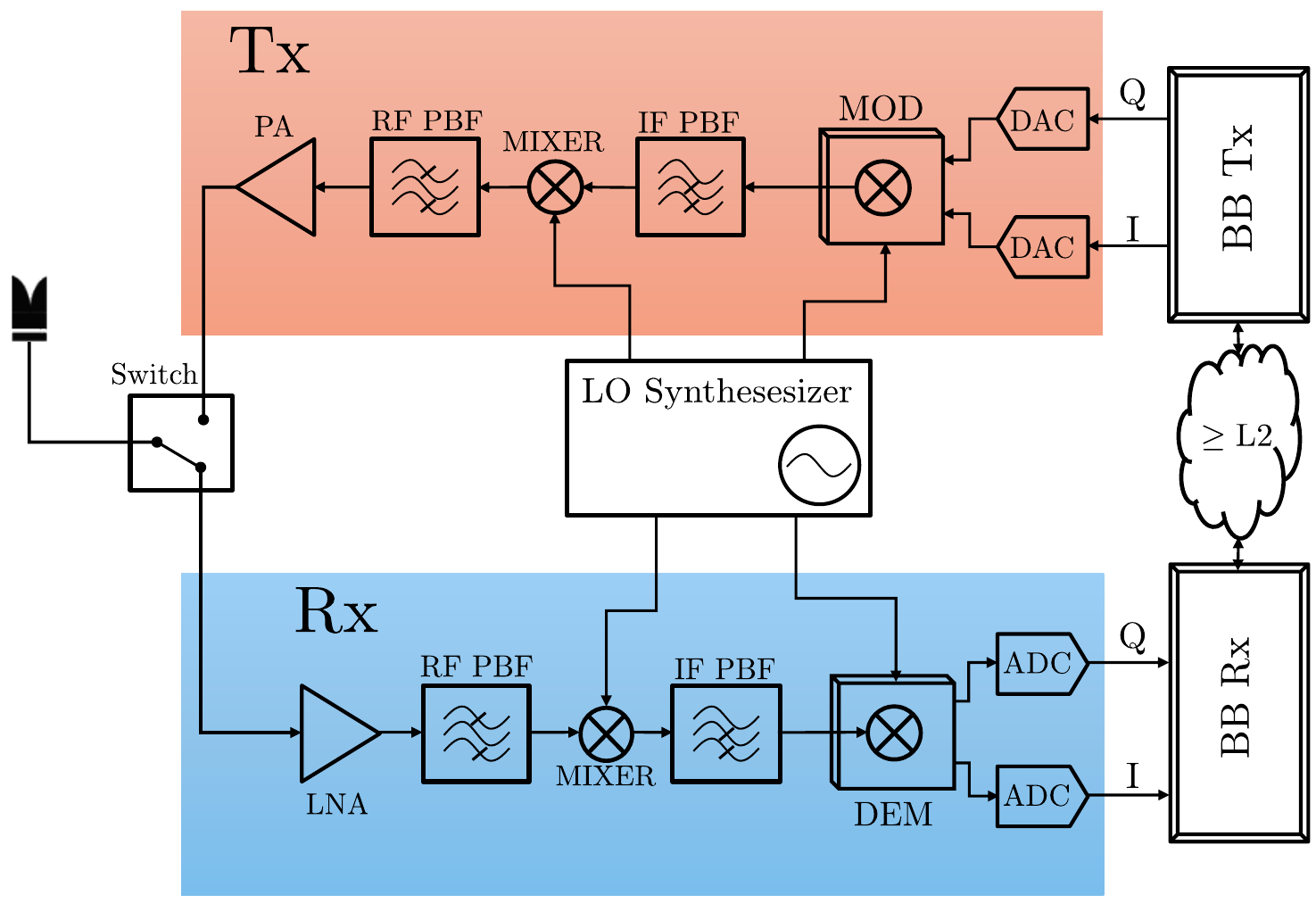}
    \caption{Generic superheterodyne wireless transceiver}
    \label{fig:transceiver}
\end{figure}

\subsection{Overview of the RF Transceiver Components}
A typical full-digital superheterodyne transceiver architecture for a wireless communication system is depicted in Fig. \ref{fig:transceiver}.
%
%
HAD architectures utilize additional components for efficient signal routing and processing, notably, phase shifters, power dividers and combiners.

In the following, we analyze the hardware impairments and energy consumption of the RF components, providing models that relate these metrics to the architectural choices for FR3 systems in Section \ref{sec:architecture}. A summary of the models used is reported in Table \ref{tab:component_summary}.

\subsubsection{Low Noise Amplifier (LNA)~\cite{arshad2013wideband}}

The power consumed by an LNA consists of static power consumption, due to bias currents $ (P_\mathrm{LNA}^s)$, and dynamic power consumption $(\kappa_\mathrm{LNA} \, B)$, which is signal-dependent:
\begin{equation}
P_\mathrm{LNA}(B) = P_\mathrm{LNA}^s + \kappa_\mathrm{LNA} \, B.
\end{equation}
The typical power consumption of LNAs ranges from 1 to 50 mW\cite{arshad2013wideband}.
The NF of an LNA is typically represented as:
\begin{equation}
\mathrm{NF}_\mathrm{LNA} = 1 + \xi_\mathrm{LNA}
\end{equation}
where \( \xi_\mathrm{LNA} \) represents the excess noise factor that depends on the design and technology. For wideband signals, a key requirement is a flat gain response, ensuring consistent amplification across the desired frequency band, without a group delay distortion. Modern LNAs exhibit an $\mathrm{NF}_\mathrm{LNA}$ range from 0.5 to 3 dB and gain between 10 to 30 dB, making them suitable for diverse communication requirements \cite{arshad2013wideband}.

\subsubsection{RF and IF Filters}

RF and IF filters are essential to isolate the desired frequency band and suppress unwanted signals, such as noise, out-of-band interference. Filters are characterized by their insertion loss (IL) and quality factor (\( Q \)). The IL is defined as:
\begin{equation}
\mathrm{IL}_{\text{F}} = 10 \log_{10}\left(\frac{P_{\text{input}}}{P_{\text{output}}}\right),
\end{equation}
and directly influences the system’s noise performance. The $Q$-factor indicates the filter's selectivity. To minimize distortion of wideband signals, RF filters require low group delay variations and flat-passband responses.

The NF of a passive filter incorporates its insertion loss along with additional contributions from thermal noise and other internal sources: 
\begin{equation}
\mathrm{NF}_\mathrm{F} = 1 + \mathrm{IL}_{\text{F}} + \xi_\mathrm{F}.
\end{equation}
where \( \xi_\mathrm{F} \) accounts for non-idealities. Passive RF filters typically exhibit insertion losses of 0.5–3 dB, while IF filters show losses between 1 and 3 dB, with \(Q\)-factors ranging from 30 to 100. 

\subsubsection{Mixers~\cite{mehta2022recent}}

Mixers shift the received signal to intermediate (IF) or baseband (BB) frequencies by combining it with a local oscillator (LO) signal. Their performance hinges on high linearity to suppress intermodulation distortion and strong isolation between RF, LO, and IF ports to minimize signal leakage.
The total power consumption of a mixer is determined by its topology and operating conditions:
\begin{equation}
P_\mathrm{mix}(B) = P^\mathrm{s}_\mathrm{mix} + \kappa_\mathrm{mix} \, B,
\end{equation}
where \(P^\mathrm{s}_\mathrm{mix}\) is the constant power due to biasing, and \(\kappa_\mathrm{mix}\) is a technology-dependent parameter. The NF is:
\begin{equation}
\mathrm{NF}_\mathrm{mix} = 1 + \xi_\mathrm{mix}.
\end{equation}
depends on \( \xi_\mathrm{mix} \) that encapsulates design-specific noise sources. Active mixers typically consume from 4 to 160 mW and have noise figures between 5 and 20 dB \cite{mehta2022recent}.

\begin{table*}[h]
\centering
\caption{Summary of Component Characteristics: Power Dissipation and Noise Figure}
\renewcommand{\arraystretch}{1.5} 
\setlength{\tabcolsep}{4pt} 
\footnotesize
\begin{tabular}{|l|>{\centering\arraybackslash}p{3cm}|>{\centering\arraybackslash}p{3cm}|p{4cm}|p{4cm}|}
\hline
\rowcolor{gray!20} 
\textbf{Component} & \textbf{\(P_\text{c}\) Model} & \textbf{\( \mathrm{NF} \) Model} & \textbf{\(P_\text{c}\) Range} & \textbf{\( \mathrm{NF} \) Range} \\ \hline
\textbf{LNA} 
& \(P_\mathrm{LNA}^s + \kappa_\mathrm{LNA} \, B\) 
& \(1 + \xi_\mathrm{LNA}\) 
& 5--50 mW \cite{arshad2013wideband}
& 0.5--6 dB  \cite{arshad2013wideband} \\ \hline
\textbf{RF/IF Filters} 
& Negligible (passive)
& \(1 + \mathrm{IL}_\text{F} + \xi_\text{F}\) 
& Negligible (passive); 10–200 mW (active) \cite{8701782} 
& 2--6 dB (passive) \cite{8701782}; 3–10 dB (active) \cite{yang2023advanced} \\ \hline
\textbf{Mixer} 
& \(P^\mathrm{s}_\mathrm{mix} + \kappa_\mathrm{mix} \, B\) 
& \(1 + \xi_\mathrm{mix}\) 
& 4–160 mW \cite{mehta2022recent} 
& 5–20 dB  \cite{mehta2022recent} \\ \hline
\textbf{ADC} 
& \(\kappa_\mathrm{ADC} \cdot B^\nu_\mathrm{ADC} \cdot 2^{N_\text{bits}}\) 
& $1 + \mathrm{Q}_\text{N} + \xi_\mathrm{ADC}$
& $>1$ mW \cite{761034} 
& $>1$ dB \cite{9674740} \\ \hline
\textbf{Divider/Combiner} 
& Negligible (passive) 
& \(1 + \mathrm{IL}_\text{D/C} + \xi_\text{D/C}\) 
& Negligible (passive) \cite{8344759}; 
& 0.1–2 dB (passive) \cite{moloudian2023design}  \\ \hline
\textbf{Phase Shifter} 
& Negligible (passive)
& \(1 + P_\mathrm{in} \,\sigma^2_\phi + L_\mathrm{PS}\) 
& Negligible (passive)
& 1–10 dB  \cite{7779294,7411411} \\ \hline
\end{tabular}
\label{tab:component_summary}
\end{table*}

\subsubsection{ADC\cite{6043594}}

The Analog-to-Digital Converter (ADC) digitizes incoming IF signals for BB processing. Power consumption is dictated by bandwidth (\(B\)), resolution (\(N_\text{bits}\)), and circuit characteristics \cite{4588351,6043594,murmann2015race}. A general model for ADC power consumption is:
\begin{equation}\label{eq:adc_power}
P_\text{ADC}(B) = \kappa_\mathrm{ADC} \cdot (2B)^\nu \cdot 2^{N_\text{bits}},
\end{equation}
where \(\kappa_\mathrm{ADC}\) is the inverse of Walden's figure-of-merit for evaluating ADC's power efficiency with resolution and speed \cite{761034,murmann2015race}, \(\nu_\mathrm{ADC}\) characterizes the scaling behavior (typical values are between 0.5 and 2), and $N_\text{bits}$ is the number of bits associated with each input sample. 

The NF of an ADC is affected by quantization noise, sampling jitter noise, comparator noise, and thermal noise stemming from the circuits' imperfections \cite{9761973}. NF can be modelled as
\begin{equation}\label{eq:adc_nf}
\mathrm{NF}_{\text{ADC}} = 1 + \mathrm{Q}_\mathrm{N} + \xi_\mathrm{ADC},
\end{equation}
where $\xi_\mathrm{ADC}$ is a technology-dependent constant, which models the sampling jitter, comparator, thermal noise, while $\mathrm{Q}_\mathrm{N}$ denotes the quantization noise, which, in dB, is defined as
\begin{equation}
  \mathrm{Q}_\mathrm{N} = -6.02 N_\text{bits} - 4.76 + \chi - 10\log_{10}\left(\frac{f_s}{2B}\right),
\end{equation}
where \(\chi\) represents the ratio between the ADC clipping level and the signal average power in dB, and \(f_s/2B\) denotes the oversampling gain. Optimizing these parameters is essential for balancing power and noise performance, especially in wideband FR3 systems. The power consumption of an ADC can differ significantly according to its features and the technology used, ranging from 1 mW to several watts \cite{761034}. This variability also applies to the NF.

\subsubsection{Signal Dividers and Combiners~\cite{8344759}}

Signal dividers and combiners are essential in HAD architectures for distributing or merging signals. Their characteristics depend on whether they are passive or active. The passive divider/combiner can be characterized by (\(\mathrm{IL}_{\text{D/C}}\)), modeled as  
\begin{equation}\label{eq:insertion_loss}
    \mathrm{IL}_{\text{D/C}} = 10 \log_{10}(N_p) + \Delta_\text{D/C},
\end{equation}
where \(N_p\) is the number of paths, and \(\Delta_\text{D/C}\) accounts for non-idealities. The noise figure is approximately
\begin{equation}\label{eq:divider_combiner_nf}
    \mathrm{NF}_{\text{D/C}} = 1 + \mathrm{IL}_{\text{D/C}} + \xi_\text{D/C}.
\end{equation}
For passive implementations, \(\mathrm{IL}_{\text{D/C}}\) and \( \mathrm{NF}_\text{D/C}\) can be very low (\(<3\,\text{dB}\)) \cite{8344759}. Active divider/combiner includes amplification stages, with power consumption between 20–160 mW, and noise figure (\(>5\,\text{dB}\)) \cite{7748489}. These designs are used when signal amplification is necessary, trading higher noise and power dissipation for improved signal strength.

\subsubsection{Phase Shifter~\cite{7779294}}

Phase shifters are essential in HAD architectures, with power consumption influenced by the implementation technology and the use of active or passive components. Digitally controlled phase shifters (DCPS), often based on switching architectures, are widely adopted in modern wireless systems due to their scalability, energy efficiency, and seamless integration with digital control systems.
The phase accuracy and noise figure (\(\mathrm{NF}_{\text{PS}}\)) of DCPS are influenced by phase quantization and circuit non-idealities. Quantization error, particularly significant in DCPS due to limited phase resolution, impacts performance. The NF is expressed as:  
\begin{equation}\label{eq:ps_nf}
\mathrm{NF}_\mathrm{PS} = 1 + P_\mathrm{in} \cdot \sigma_\phi^2 + \mathrm{IL}_\mathrm{PS},
\end{equation}
where \(P_\mathrm{in}\) is the input signal power, \(\sigma_\phi^2\) denotes the phase error variance due to quantization, $\mathrm{IL}_\mathrm{PS}$ is the loss. The variance \(\sigma_\phi^2\) decreases with increased resolution. Typical DCPS NF ranges between 1–10 dB \cite{7779294,7411411}. In wideband systems, PS limitations can lead to beam squinting, that is, a frequency-dependent deviation of the beam direction. Techniques such as true-time delay networks and hybrid analog-digital architectures are often employed to address these challenges and ensure consistent performance (not considered herein, see \cite{9771341}).

\section{Hardware Architectures for FR3 Band Systems}\label{sec:architecture}

Based on the schematic noise and power consumption models in Sect. III, here the architectural design choices of the RF chain and the processing strategies that support efficient communication systems at FR3. We propose to divide the architectures for MIMO multi-band FR3 into two primary classes: \textit{Frequency Integrated (FI)} and \textit{Frequency Partitioned (FP)}. Each category is analyzed herein with respect to the required components, system complexity, highlighting the trade-offs in terms of energy efficiency and performance.

\subsection{Frequency-Integrated (FI) Architectures}

\begin{figure*}
    \centering
    \subfloat[FD SRF]{\includegraphics[width=0.45\linewidth]{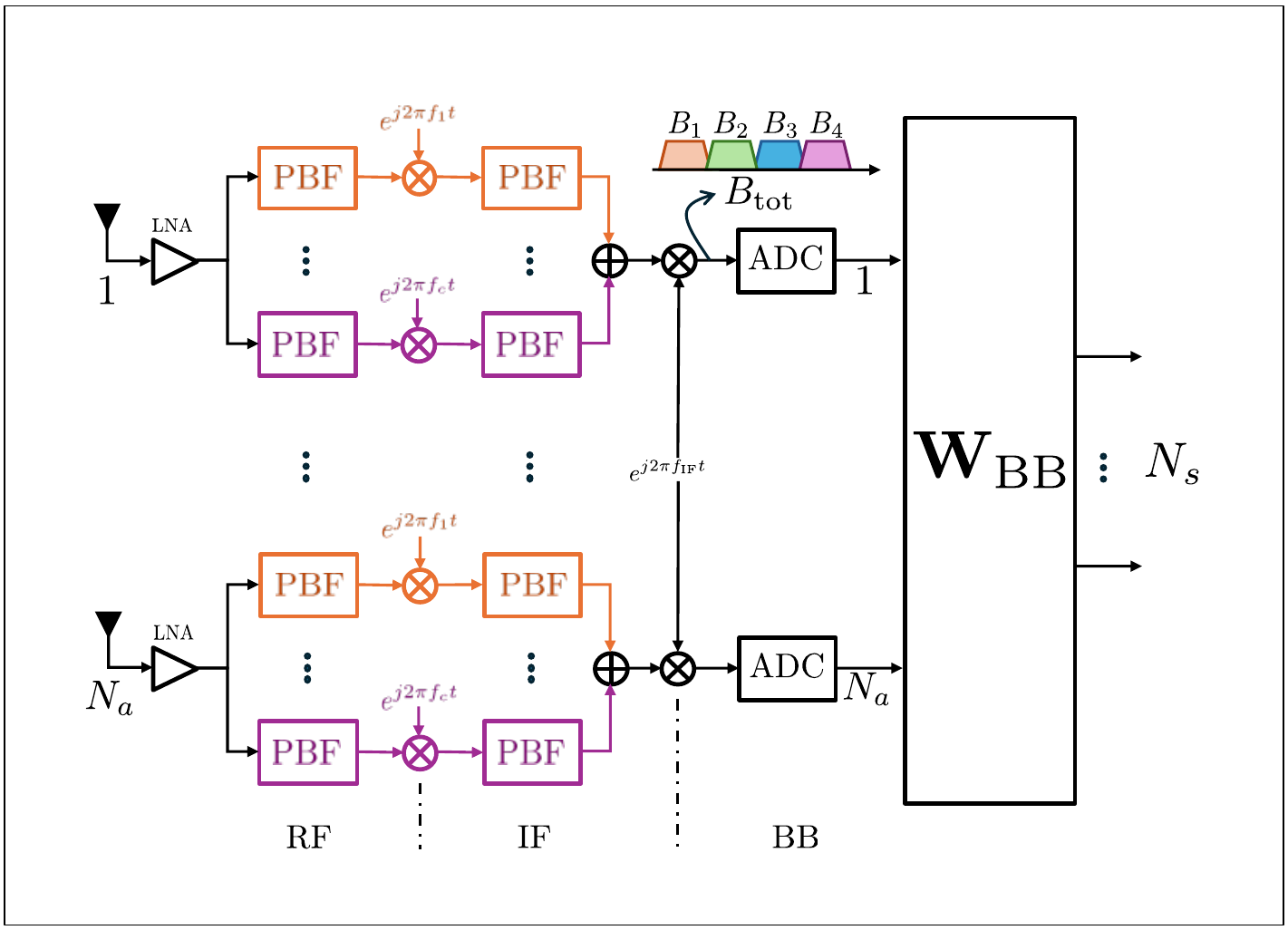}\label{subfig:FIFDS}} \hspace{.1cm}
    \subfloat[FD DRF]{\includegraphics[width=0.45\linewidth]{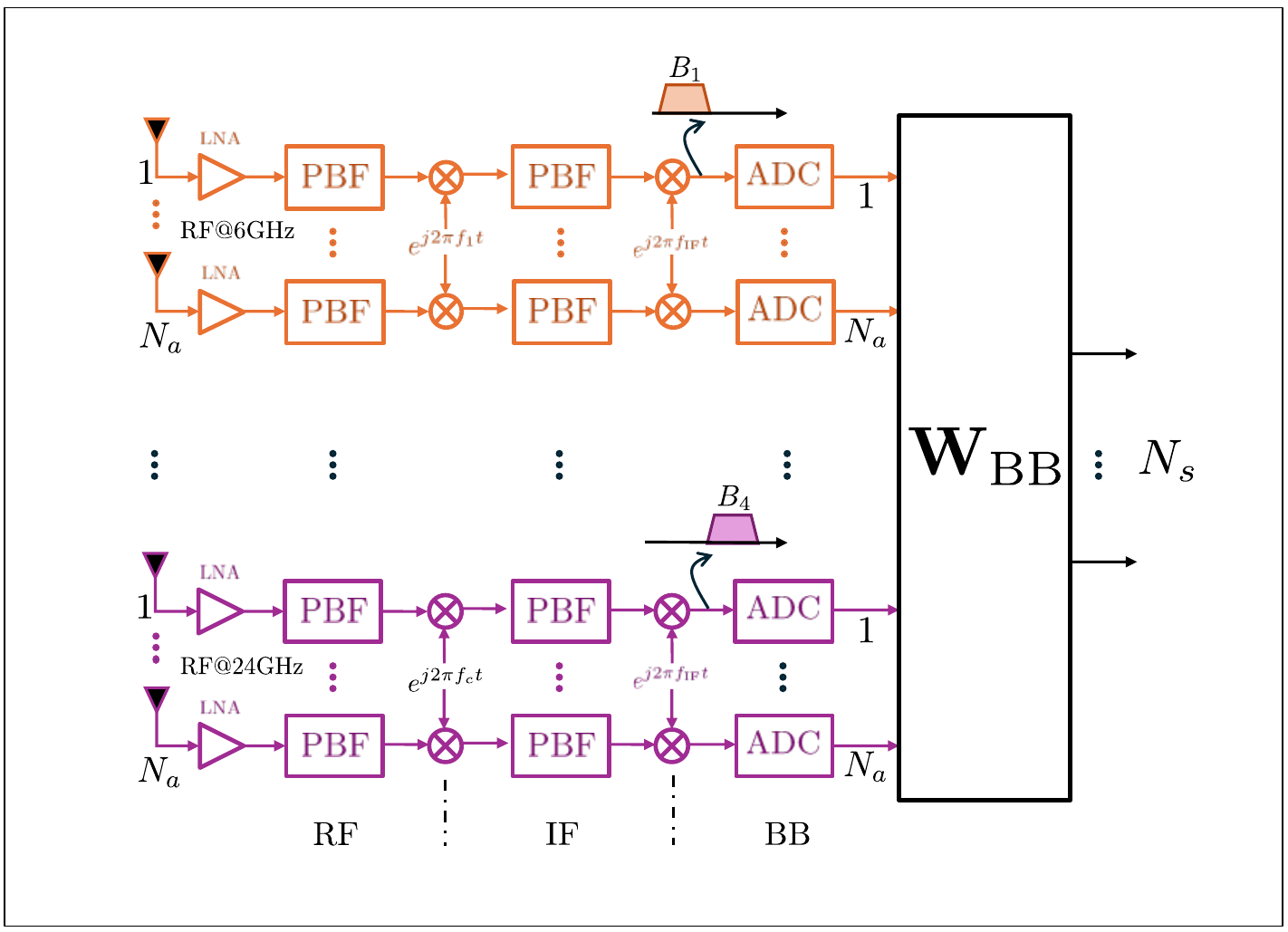}\label{subfig:FIFDNS}} \\
    \subfloat[HAD SRF]{\includegraphics[width=0.45\linewidth]{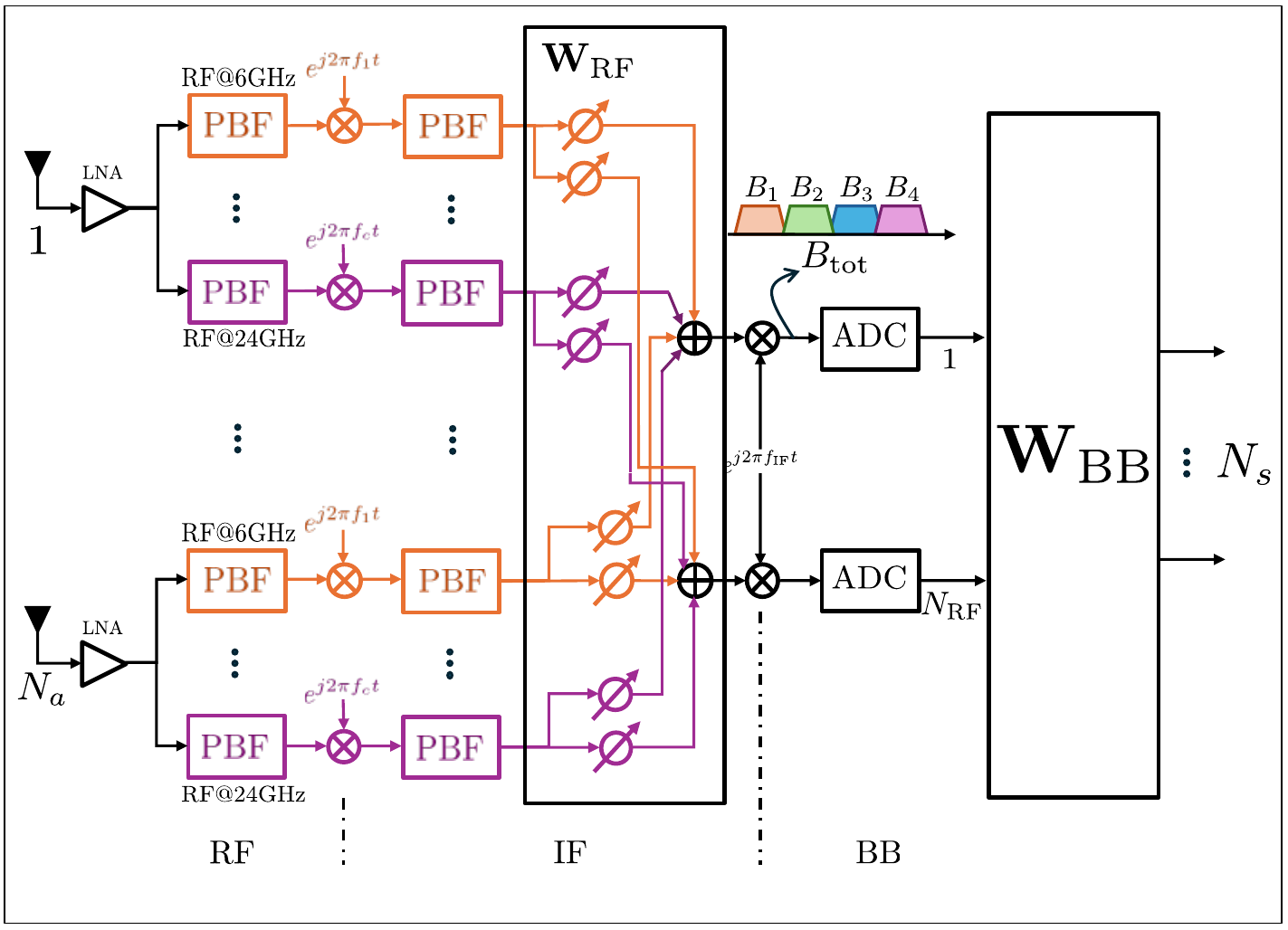}\label{subfig:FIHADS}} \hspace{0.1cm}
    \subfloat[HAD DRF]{\includegraphics[width=0.45\linewidth]{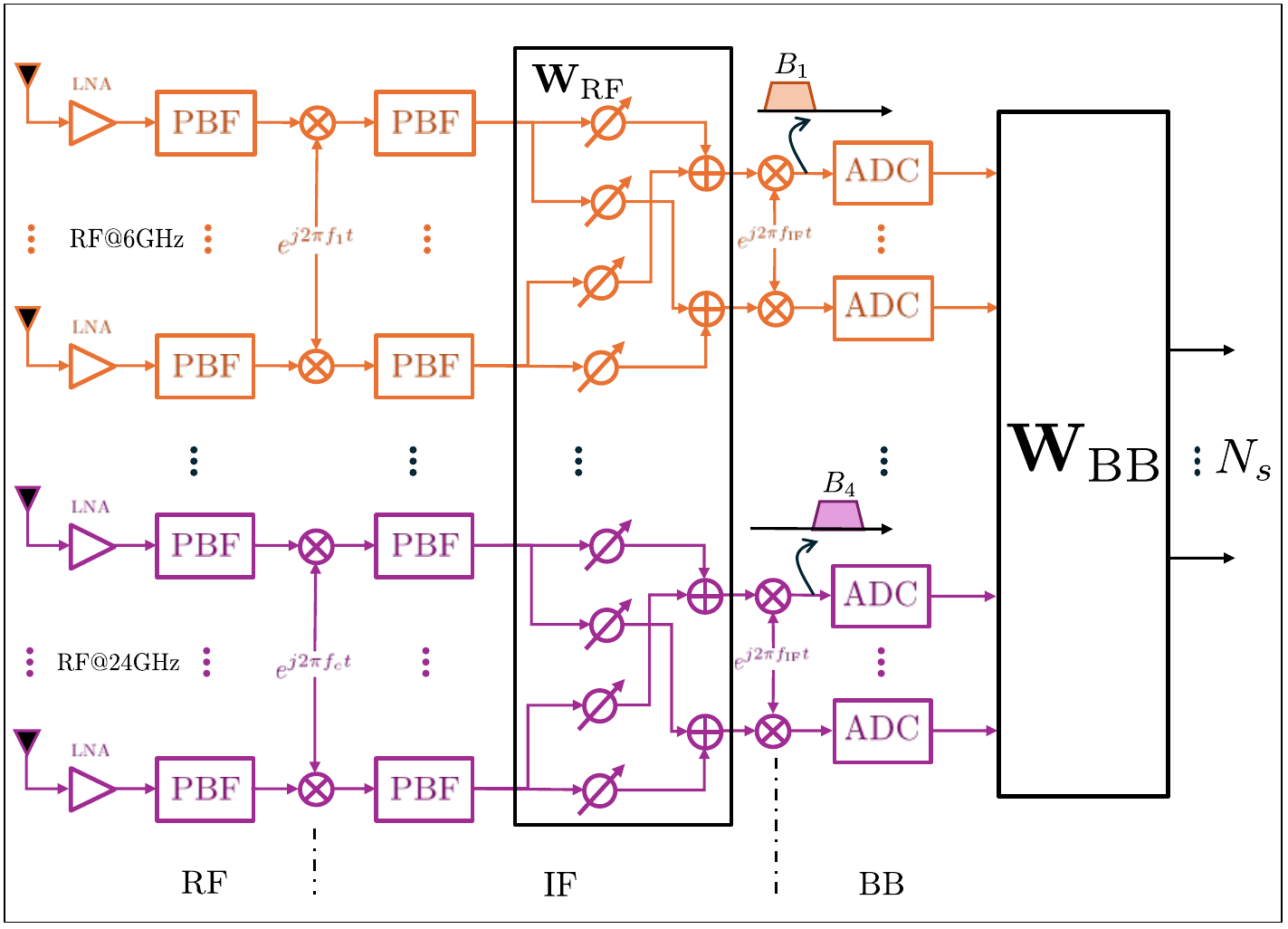}\label{subfig:FIHADNS}} 
    
    \caption{Frequency Integrated (FI) Architectures: Enables simultaneous access to all sub-bands.}
    \label{fig:FI}
\end{figure*}

FI architectures, illustrated in Fig. \ref{fig:FI}, enable simultaneous access to all sub-bands by heterodyning\footnote{Frequency down/up conversion of multiple frequencies simultaneously.} multiple carriers, thereby maximizing spectral efficiency and system capacity. However, this approach introduces significant challenges in both signal processing and circuit integration due to the increased complexity of multi-band operation. In FI designs, RF chains can either be shared among all sub-bands or allocated separately to each sub-band, while baseband processing is always performed jointly across all the sub-bands to achieve optimal performance.

Fewer components are needed when the RF chains are shared, simplifying the hardware design. However, shared components, such as antennas and LNAs, must span the full RF spectrum, while ADCs and DACs must also operate over the total bandwidth of all sub-bands ($B_{\text{tot}} = \sum_c B_c$), creating significant challenges as current technology struggles to achieve high performance at such wide bandwidths \cite{kang2024cellular}.
By contrast, assigning dedicated RF chains and components to each sub-band increases the number of components but relaxes bandwidth requirements on individual parts. This makes the components easier to manufacture, as they only handle the bandwidth of a single sub-band rather than the entire spectrum.

\subsubsection{FI FD with Shared RF Chains}

This architecture, depicted in Fig. \ref{subfig:FIFDS}, shares RF chains and antennas across all sub-bands. Digital baseband processing $\mathbf{W}_\mathrm{BB}[c]$ handles the combined bandwidth of all sub-bands ($B_{\text{tot}} = \sum_c B_c$), ensuring flexibility and full spatial processing capabilities. Its processing structure, power consumption, and noise are listed below:

\begin{itemize}
    \item beamforming Matrix:
    \begin{equation}
        \mathbf{W}[c] = \mathbf{W}_{\text{BB}}[c]
    \end{equation}
    \item Power Consumption: 
    \begin{equation}
        P_T = P_S + P_D(B)
    \end{equation}
    The static power $P_S$ consumption remains constant regardless of system activity and is primarily determined by the number of active components. In contrast, dynamic power $P_D(B)$ consumption occurs when a signal, with a specific bandwidth, passes through the component. The static power consumption \( P_s \) is given by:  
    \begin{equation}
    \begin{split}  
        P_S \propto N_a &\bigg( P_\mathrm{LNA}^\mathrm{s} + 4 P^\mathrm{RF-IF,s}_\mathrm{mix} \;+\\
        &+ P^\mathrm{IF-BB,s}_\mathrm{mix} +2 P_\mathrm{ADC}^\mathrm{s} \bigg),
    \end{split}  
    \end{equation}
    and the dynamic power consumption \( P_d \) is  
    \begin{equation}
    \begin{split}  
        P_D(B) \propto N_a \,B_T &\bigg[ \big( \kappa_\mathrm{LNA} + \kappa_\mathrm{mix}^\mathrm{RF-IF} + \kappa_\mathrm{mix}^\mathrm{IF-BB}\big) + \\
        &+ 2^{\mathrm{N_{bits}}+\nu} \, \kappa_\mathrm{ADC} B_T^{\nu-1} \bigg]
    \end{split}  
    \end{equation}
    \item Noise Figure: is calculated using \eqref{eq:friis} with the cascade of components: LNA, signal divider, RF Filter, RF-IF Mixer, IF Filter, IF-BB Mixer, signal combiner, and ADC. It is important to note that since signals at different carriers remain distinct in IF frequency and do not merge, each carrier would have the same NF since identical components are systematically applied to all carriers.
\end{itemize}

\subsubsection{FI HAD with Shared RF Chains}

This architecture, depicted in Fig. \ref{subfig:FIHADS}, reduces the number of RF chains ($N_{\text{RF}} < N_a$) using a hybrid approach. The RF chains are connected to all antennas (fully connected) or to a subset of antennas (subconnected), with analog beamforming by phase shifters performed before demodulation.

\begin{itemize}
    \item beamforming Matrix: 
    \begin{equation}
        \mathbf{W}[c] = \mathbf{W}_{\text{RF}}[c] \mathbf{W}_{\text{BB}}[c]
    \end{equation}
    where $\mathbf{W}_{\text{RF}}[c] \in \mathbb{C}^{N_a \times N_{\text{RF}}}$ and $\mathbf{W}_{\text{BB}}[c] \in \mathbb{C}^{N_{\text{RF}} \times K_c}$ are the analog and digital beamforming matrices. Since \( \mathbf{W}_\mathrm{RF} \) is implemented using analog phase shifters, its elements are constrained to satisfy  $(\mathbf{W}_\mathrm{RF}^{(i)} \mathbf{W}_\mathrm{RF}^{\mathrm{H}(i)})_{\ell,\ell} = 1/N_a$,
    where \( (\cdot)_{\ell,\ell} \) denotes the $\ell$th diagonal element of a matrix, and the pedix $(i)$ refer to the $i$th column of  \( \mathbf{W}_\mathrm{RF} \).
    \item Power Consumption:
    the static power consumption \( P_s \) is given by:  
    \begin{equation}
    \begin{split}  
        P_s &\propto N_a \bigg( P_\mathrm{LNA}^\mathrm{s} + 4 P^\mathrm{RF-IF,s}_\mathrm{mix} \bigg) \\
        &+ N_\mathrm{RF}\bigg( P^\mathrm{IF-BB,s}_\mathrm{mix} +2 P_\mathrm{ADC}^\mathrm{s} \bigg),
    \end{split}  
    \end{equation}
    and the dynamic power consumption \( P_d \) is  
    \begin{equation}
    \begin{split}  
        P_D(B) \propto B_T &\bigg[ N_a \big( \kappa_\mathrm{LNA} + \kappa_\mathrm{mix}^\mathrm{RF-IF}\big) \;+ \\
        &+ N_\mathrm{RF} \big( \kappa_\mathrm{mix}^\mathrm{IF-BB} + 2^{\mathrm{N_{bits}}+\nu} \kappa_\mathrm{ADC} B_T^{\nu-1} \big) \bigg]
    \end{split}  
    \end{equation}
    \item Noise Figure: is calculated using \eqref{eq:friis} with the cascade of components: LNA, signal divider, RF filter, RF-IF mixer, IF filter, phase-shifter, signal combiner, IF-BB mixer, and ADC. 
\end{itemize}

Differently from conventional HAD MIMO systems \cite{6717211}, the analog beamforming matrix depends on the considered sub-band $\mathbf{W}_\mathrm{RF}[c]$. This is because currently designed phase shifters are frequency selective, and using the same phase shifter for all sub-bands will result in beam-squinting effects (or beam-splitting) \cite{wan2021hybrid}. Therefore, for FI architectures, it is recommended that a dedicated phase-shifting network is used for each sub-band.

\subsubsection{FI FD with Dedicated RF Chains}

In this architecture, depicted in Fig. \ref{subfig:FIFDNS}, each sub-band uses dedicated RF chains and antennas, providing isolation and independent operation for each sub-band.

\begin{itemize}
    \item beamforming Matrix:
    \begin{equation}
    \mathbf{W} = \mathrm{Blkdiag}\bigg(\mathbf{W}_{\text{BB}}[1],\mathbf{W}_{\text{BB}}[2],\mathbf{W}_{\text{BB}}[3],\mathbf{W}_{\text{BB}}[4]\bigg)
    \end{equation}
    
    \item Power Consumption:
    the static power consumption \( P_s \) is given by:  
    \begin{equation}
    \begin{split}  
        P_s \propto 4\,N_a &\bigg( P_\mathrm{LNA}^\mathrm{s} + P^\mathrm{RF-IF,s}_\mathrm{mix} \;+\\
        &+ P^\mathrm{IF-BB,s}_\mathrm{mix} +2 P_\mathrm{ADC}^\mathrm{s} \bigg),
    \end{split}  
    \end{equation}
    and the dynamic power consumption \( P_d \) is  
    \begin{equation}
    \begin{split}  
        P_D(B) \propto N_a\,\sum_c  B_c &\bigg[ \big( \kappa_\mathrm{LNA} + \kappa_\mathrm{mix}^\mathrm{RF-IF} + \kappa_\mathrm{mix}^\mathrm{IF-BB} \big) +\\ &+2^{\mathrm{N_{bits}}+\nu} \, \kappa_\mathrm{ADC}   \,B_c^{\nu-1} \bigg]
    \end{split}  
    \end{equation}
    \item Noise Figure: is calculated using \eqref{eq:friis} with the cascade of components: LNA, RF Filter, RF-IF Mixer, IF Filter, IF-BB Mixer, and ADC. 
\end{itemize}

\begin{figure*}
    \centering
    \subfloat[FD SRF]{\includegraphics[width=0.45\linewidth]{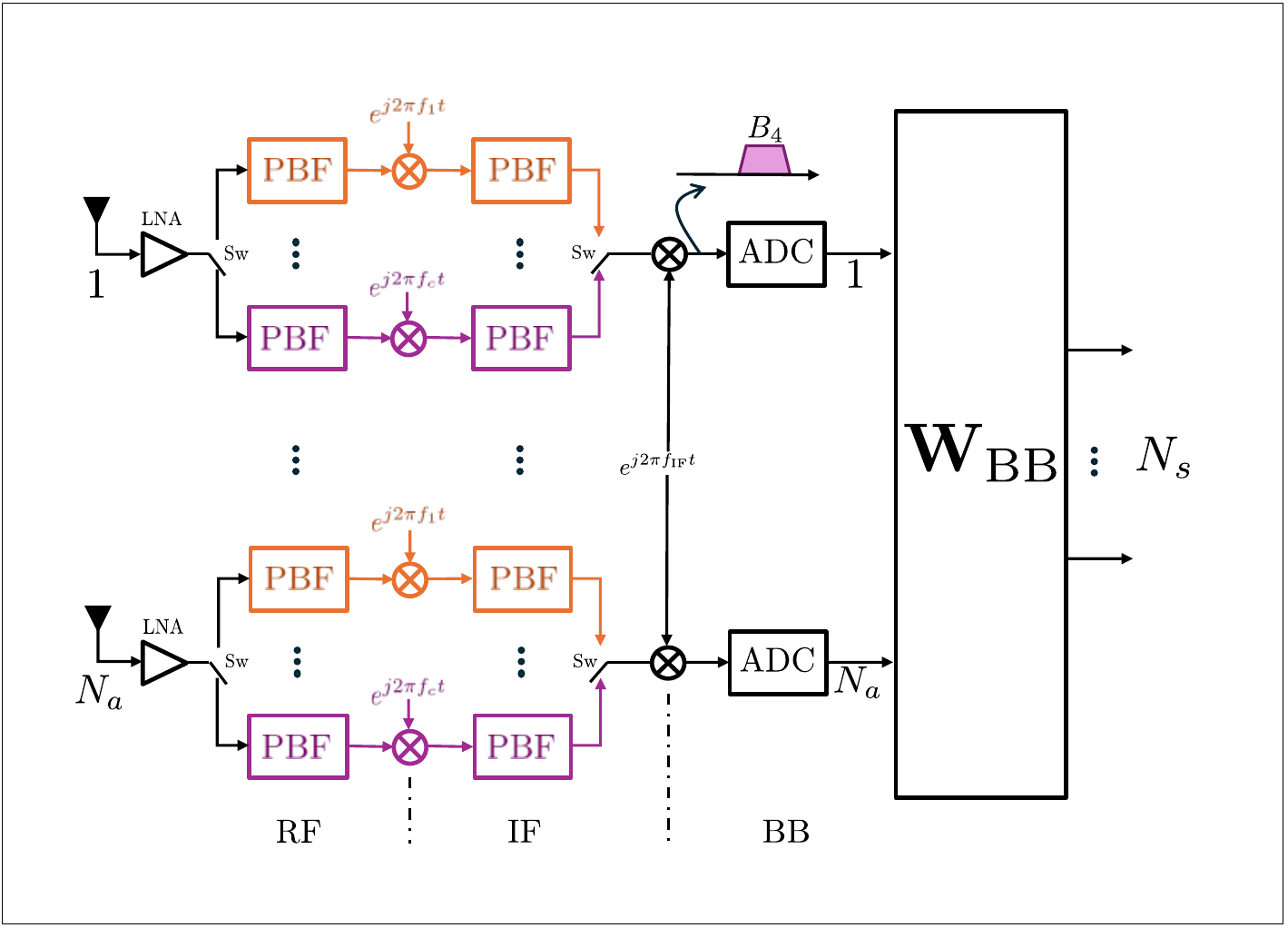}\label{subfig:FPFDS}} \hspace{.1cm}
    \subfloat[FD DRF]{\includegraphics[width=0.45\linewidth]{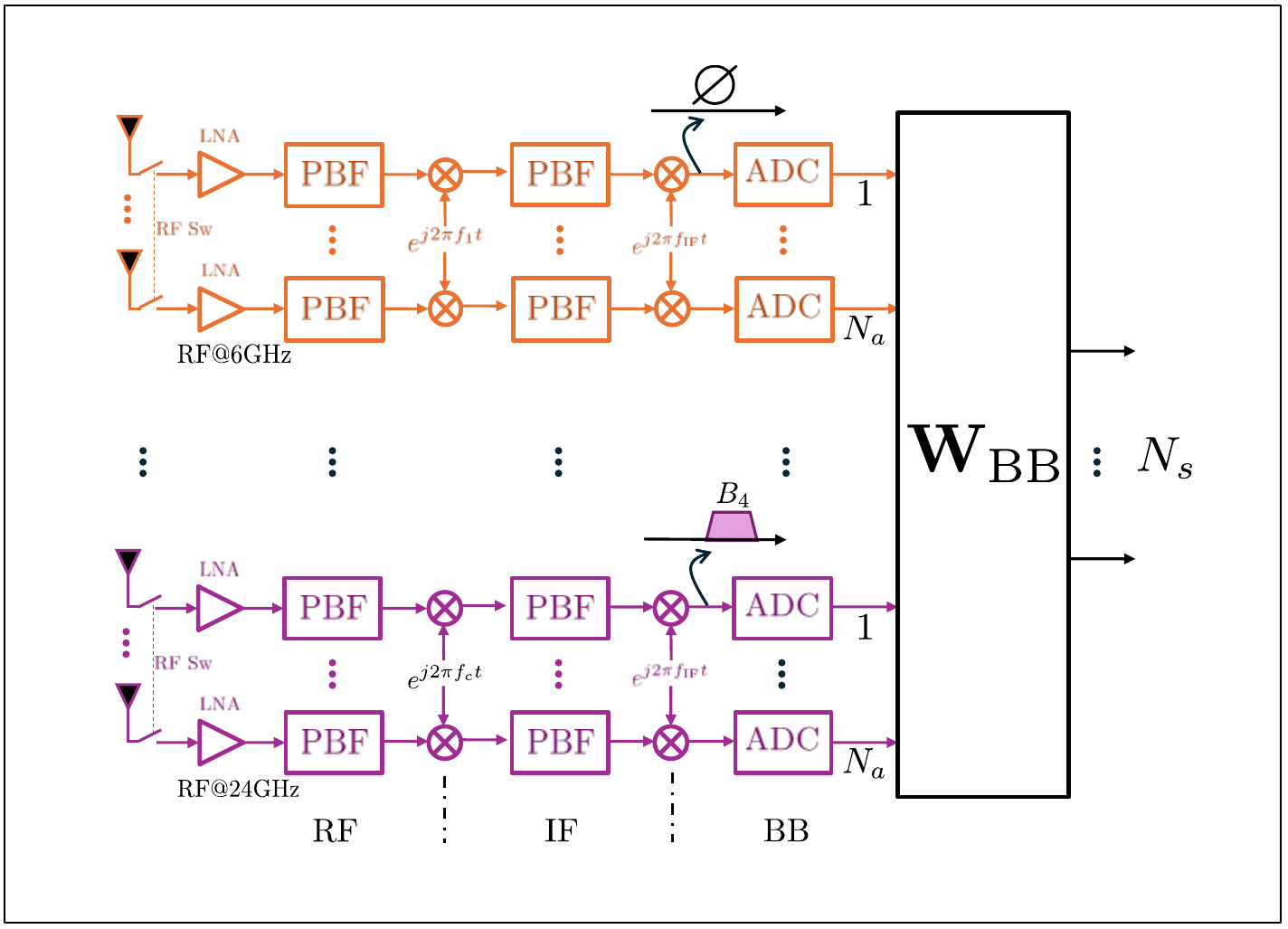}\label{subfig:FPFDNS}} \\
    \subfloat[HAD SRF]{\includegraphics[width=0.45\linewidth]{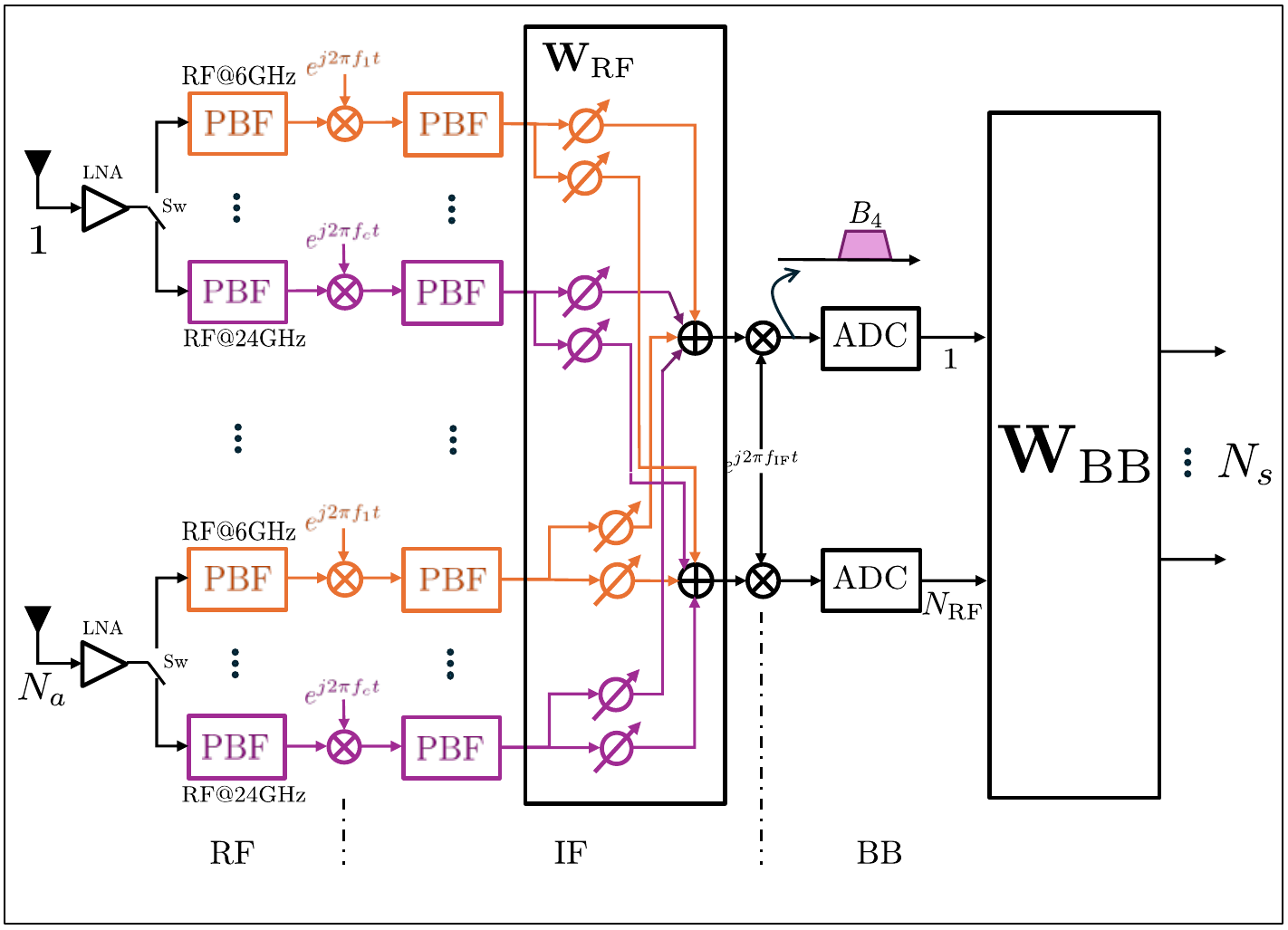}\label{subfig:FPHADS}} \hspace{.1cm}
    \subfloat[HAD DRF]{\includegraphics[width=0.45\linewidth]{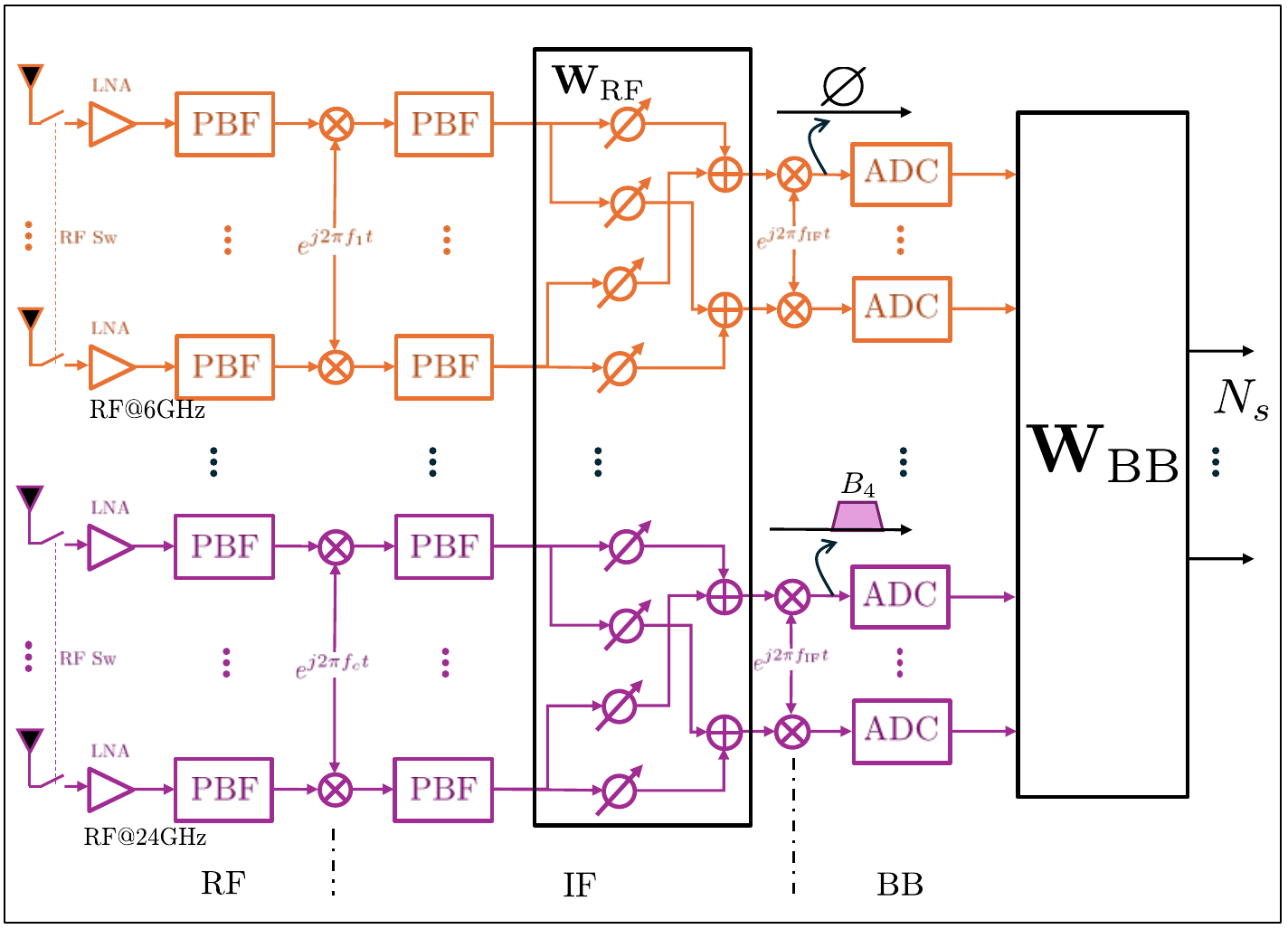}\label{subfig:FPHADNS}} 
    
    \caption{Frequency-Partitioned (FP) Architectures: A Switch after the LNAs selects the sub-band to serve.}
    \label{fig:FP}
\end{figure*}

\subsubsection{FI HAD with Dedicated RF Chains}

This architecture, depicted in Fig. \ref{subfig:FIHADNS}, combines analog and digital beamforming with independent RF chains per sub-band. Both analog and digital beamforming matrices are block diagonal.

\begin{itemize}
    \item beamforming Matrices:
    \begin{align}
    \mathbf{W}_{\text{RF}} &= \mathrm{Blkdiag}\bigg(\mathbf{W}_{\text{RF}}[1],\mathbf{W}_{\text{RF}}[2],\mathbf{W}_{\text{RF}}[3],\mathbf{W}_{\text{RF}}[4]\bigg)
    \\
    \mathbf{W}_{\text{BB}} &= \mathrm{Blkdiag}\bigg(\mathbf{W}_{\text{BB}}[1],\mathbf{W}_{\text{BB}}[2],\mathbf{W}_{\text{BB}}[3],\mathbf{W}_{\text{BB}}[4]\bigg)
    \end{align}
    \item Power consumption:
    the static power consumption \( P_s \) is given by:  
    \begin{equation}
    \begin{split}  
        P_s &\propto 4\, N_a \bigg( P_\mathrm{LNA}^\mathrm{s} +  P^\mathrm{RF-IF,s}_\mathrm{mix} \bigg) \\
        &+ 4\,N_\mathrm{RF}\bigg( P^\mathrm{IF-BB,s}_\mathrm{mix} +2 P_\mathrm{ADC}^\mathrm{s} \bigg),
    \end{split}  
    \end{equation}
    and the dynamic power consumption \( P_d \) is  
    \begin{equation}
    \begin{split}  
        P_D(B) \propto &\sum_c  B_c \bigg[ N_a\,\big( \kappa_\mathrm{LNA} + \kappa_\mathrm{mix}^\mathrm{RF-IF} \big) + \\
        &+ N_\mathrm{RF} \big(\kappa_\mathrm{mix}^\mathrm{IF-BB}  +2^{\mathrm{N_{bits}}+\nu} \, \kappa_\mathrm{ADC}   \,B_c^{\nu-1} \big)\bigg]
    \end{split}  
    \end{equation}
    \item Noise Figure: is calculated using \eqref{eq:friis} with the cascade of components: LNA, RF filter, RF-IF mixer, IF filter, signal divider,  phase-shifter, signal combiner, IF-BB mixer, and ADC. 
\end{itemize}

\subsection{Frequency-Partitioned (FP) Architectures}

\begin{table*}[t]
\centering
\caption{Summary of FI and FP Architectures with Component Counts}
\renewcommand{\arraystretch}{1.5} 
\setlength{\tabcolsep}{2pt} 
\begin{tabular}{|c|l|c|c|c|c|c|c|}
\hline
\rowcolor{gray!20} 
\textbf{Class} & \textbf{Architecture} & \textbf{Antennas} & \textbf{ADCs} & \textbf{Phase Shifters} & \textbf{RF/IF Filter} & \textbf{IF-BB Mixer} & \textbf{Divider/Combiner} \\ \hline \hline
\multirow{4}{*}{\textbf{FI}} 
& \textbf{FD SRF} & $N_a$ & $2\times N_a$ & 0 & $4\times N_a$ & $5\times N_a$ & $ N_a/N_a$ \\ \cline{2-8}
& \textbf{HAD SRF} & $N_a$ & $2\times N_{\text{RF}}$ & $4\times N_a N_{\text{RF}}$ & $4\times N_a$ & $4\times N_a + N_{\text{RF}}$ & $5\times N_a/N_\mathrm{RF}$ \\ \cline{2-8}
& \textbf{FD DRF} & $4 \times N_a$ & $8 \times N_a$ & 0 & $8 \times N_a$ & $8 \times N_a$ & 0 \\ \cline{2-8}
& \textbf{HAD DRF} & $4 \times N_a$ & $8 \times N_{\text{RF}}$ & $4 \times N_a N_{\text{RF}}$ & $8 \times N_a$ & $4 \times N_a + 4 \times N_{\text{RF}}$ & $4 \times N_a/4 \times N_{\text{RF}}$ \\ \hline \hline
\multirow{4}{*}{\textbf{FP}} 
& \textbf{FD SRF} & $N_a$ & $2\times N_a$ & 0 & $4\times N_a$ & $5\times N_a$ & $0$ \\ \cline{2-8}
& \textbf{HAD SRF} & $N_a$ & $2\times N_{\text{RF}}$ & $4\times N_a N_{\text{RF}}$ & $4\times N_a$ & $4\times N_a + N_{\text{RF}}$ & $4\times N_a/N_\mathrm{RF}$ \\ \cline{2-8}
& \textbf{FD DRF} & $4 \times N_a$ & $8 \times N_a$ & 0 & $8 \times N_a$ & $8 \times N_a$ & 0 \\ \cline{2-8}
& \textbf{HAD DRF} & $4 \times N_a$ & $8 \times N_{\text{RF}}$ & $4 \times N_a N_{\text{RF}}$ & $8 \times N_a$ & $4 \times N_a + N_{\text{RF}}$ & $4 \times N_a/ 4 \times N_\mathrm{RF}$ \\ \hline
\end{tabular}
\label{tab:componentComparison}
\end{table*}

FP architectures are depicted in Fig. \ref{fig:FP} and operate dynamically by accessing one sub-band at a time in a time-division manner. This approach simplifies the BB processing and reduces power consumption compared to Frequency-Integrated (FI) architectures. However, the sequential nature of sub-band processing can limit the overall spectral efficiency and throughput. The FP architectures incorporate RF switching mechanisms to enable dynamic sub-band selection for operation at any given time. This flexibility allows the system to adapt to varying channel conditions and user demands while maintaining reduced hardware complexity.

When RF chains are shared across sub-bands, the total number of required components is minimized. An RF switch dynamically selects which sub-band is active, and only the selected sub-band is processed. However, the shared components, such as ADCs, must operate over the largest bandwidth among the sub-bands.
In architectures with dedicated RF chains, each sub-band is associated with a dedicated set of RF chains and components. During operation, the RF switch activates the specific RF chains associated with the sub-band in use. This approach increases the total number of components but reduces the demands on individual components since they only need to operate within the bandwidth of their corresponding sub-band. Consequently, these components can be manufactured with relative ease using current technologies.

\subsubsection{FP FD with Shared RF Chains}

This design proposed by \cite{10198042} is depicted in Fig. \ref{subfig:FPFDS}. The RF chains are shared across antennas within each sub-band. Digital baseband processing is performed independently for each sub-band, allowing for efficient resource allocation.

\begin{itemize}
    \item Combination matrix in time slot $t$:
    \begin{equation}
        \mathbf{W}[t] = \mathbf{W}_{\text{BB}}[c]
    \end{equation}
    \item Power Consumption:
    the static power consumption \( P_s \) is given by:  
    \begin{equation}
    \begin{split}  
        P_s \propto N_a &\bigg( P_\mathrm{LNA}^\mathrm{s} + 4 P^\mathrm{RF-IF,s}_\mathrm{mix} \;+\\
        &+ P^\mathrm{IF-BB,s}_\mathrm{mix} +2 P_\mathrm{ADC}^\mathrm{s} \bigg)  
    \end{split}  
    \end{equation}
    The dynamic power consumption \( P_d \) is  
    \begin{equation}
    \begin{split}
        P_D(B) \propto N_a \sum_c &\bigg[B_c \big( \kappa_\mathrm{LNA} + \kappa_\mathrm{mix}^\mathrm{RF-IF} +\kappa_\mathrm{mix}^\mathrm{IF-BB} \big) + \\
        &+\bigg(2^{\mathrm{N_{bits}}+\nu} \,\kappa_\mathrm{ADC} \, B_c^{\nu-1} \bigg) \bigg] \tau_c
    \end{split}
    \end{equation}
    where $\sum_c \tau_c = 1$, with $\tau_c$ denoting the average time dedicated to sub-band $c$.
    
    \item Noise Figure: is calculated using \eqref{eq:friis} with the cascade of components: LNA, RF Filter, RF-IF Mixer, IF Filter, IF-BB Mixer, and ADC. 
\end{itemize}

\subsubsection{FP HAD with Shared RF Chains}

This architecture, depicted in Fig. \ref{subfig:FPHADS}, employs fewer RF chains per sub-band using a hybrid beamforming approach. Analog beamforming is applied to each sub-band independently before digital processing.

\begin{itemize}
    \item beamforming matrix in time slot $t$:
    \begin{equation}
        \mathbf{W}[t] = \mathbf{W}_{\text{RF}}[c]\mathbf{W}_{\text{BB}}[c]
    \end{equation}
    \item Power Consumption:
    the static power consumption \( P_s \) is given by:  
    \begin{equation}
    \begin{split}  
        P_s &\propto N_a \bigg( P_\mathrm{LNA}^\mathrm{s} + 4 P^\mathrm{RF-IF,s}_\mathrm{mix} \bigg) \\
        &+ N_\mathrm{RF}\bigg( P^\mathrm{IF-BB,s}_\mathrm{mix} +2 P_\mathrm{ADC}^\mathrm{s} \bigg)  
    \end{split}  
    \end{equation}
    The dynamic power consumption \( P_d \) is  
    \begin{equation}
    \begin{split}  
        P_D(B) \propto & \sum_c B_c\,\tau_c \bigg[ N_a  \big( \kappa_\mathrm{LNA} + \kappa_\mathrm{mix}^\mathrm{RF-IF} \big) + \\
        &+ N_\mathrm{RF} \bigg( \kappa_\mathrm{mix}^\mathrm{IF-BB} \, + 2^{\mathrm{N_{bits}}+\nu} \kappa_\mathrm{ADC}\, B_c^{\nu-1} \bigg) \bigg]
    \end{split}  
    \end{equation}
    \item Noise Figure: is calculated using \eqref{eq:friis} with the cascade of components: LNA, RF filter, RF-IF mixer, IF filter, signal divider,  phase-shifter, signal combiner, IF-BB mixer, and ADC. 
\end{itemize}

\subsubsection{FP FD with Dedicated RF Chains}

In this architecture, as depicted in Fig. \ref{subfig:FPFDNS}, each sub-band operates with independent RF chains and antennas, ensuring complete isolation and maximum per-sub-band flexibility. 

\begin{itemize}
    \item beamforming matrix in time slot $t$:
    \begin{equation}
    \mathbf{W}[t] = \mathrm{Blkdiag}\bigg(\mathbf{0},\mathbf{W}_{\text{BB}}[c],\mathbf{0},\mathbf{0}\bigg)
    \end{equation}
    \item Power Consumption:
    the static power consumption \( P_s \) is given by:  
    \begin{equation}
    \begin{split}  
        P_s \propto 4\,N_a &\bigg( P_\mathrm{LNA}^\mathrm{s} + P^\mathrm{RF-IF,s}_\mathrm{mix} \;+\\
        &+ P^\mathrm{IF-BB,s}_\mathrm{mix} +2 P_\mathrm{ADC}^\mathrm{s} \bigg),
    \end{split}  
    \end{equation}
    and the dynamic power consumption \( P_d \) is  
    \begin{equation}
    \begin{split}  
        P_D(B) \propto N_a\,\sum_c  B_c &\bigg[ \big( \kappa_\mathrm{LNA} + \kappa_\mathrm{mix}^\mathrm{RF-IF} + \kappa_\mathrm{mix}^\mathrm{IF-BB} \big) +\\ &+2^{\mathrm{N_{bits}}+\nu} \, \kappa_\mathrm{ADC}   \,B_c^{\nu-1} \bigg] \tau_c
    \end{split}  
    \end{equation}
    
    \item Noise Figure: is calculated using \eqref{eq:friis} with the cascade of components: LNA, RF Filter, RF-IF Mixer, IF Filter, IF-BB Mixer, and ADC. 
\end{itemize}

\subsubsection{FP HAD with Dedicated RF Chains}

This approach, depicted in Fig. \ref{subfig:FPHADNS}, combines analog and digital beamforming for each sub-band independently, with independent RF chains. Both analog and digital beamforming matrices are block diagonal.

\begin{itemize}
    \item beamforming matrix in time slot $T_s$, assuming sub-band $c=2$ is active:
    \begin{align}
    \mathbf{W}_{\text{RF}} &= \mathrm{Blkdiag}\bigg(\mathbf{0},\mathbf{W}_{\text{RF}}[c],\mathbf{0},\mathbf{0}\bigg)
    \\
    \mathbf{W}_{\text{BB}} &= \mathrm{Blkdiag}\bigg(\mathbf{0},\mathbf{W}_{\text{BB}}[c],\mathbf{0},\mathbf{0}\bigg)
     \end{align}
    \item Power consumption:
    the static power consumption \( P_s \) is given by:  
    \begin{equation}
    \begin{split}  
        P_s &\propto 4\, N_a \bigg( P_\mathrm{LNA}^\mathrm{s} +  P^\mathrm{RF-IF,s}_\mathrm{mix} \bigg) \\
        &+ 4\,N_\mathrm{RF}\bigg( P^\mathrm{IF-BB,s}_\mathrm{mix} +2 P_\mathrm{ADC}^\mathrm{s} \bigg),
    \end{split}  
    \end{equation}
    and the dynamic power consumption \( P_d \) is  
    \begin{equation}
    \begin{split}  
        P_D(B) \propto &\sum_c  B_c\, \tau_c \bigg[ N_a\,\big( \kappa_\mathrm{LNA} + \kappa_\mathrm{mix}^\mathrm{RF-IF} \big) + \\
        &+ N_\mathrm{RF} \big(\kappa_\mathrm{mix}^\mathrm{IF-BB}  +2^{\mathrm{N_{bits}}+\nu} \, \kappa_\mathrm{ADC}   \,B_c^{\nu-1} \big)\bigg]
    \end{split}  
    \end{equation}
    \item Noise Figure: is calculated using \eqref{eq:friis} with the cascade of components: LNA, RF filter, RF-IF mixer, IF filter, signal divider, phase-shifter, signal combiner, IF-BB mixer, and ADC. 
\end{itemize}

Table \ref{tab:componentComparison} provides an overview of the key RF components required for each architecture discussed in Sect. VI. By associating a unitary cost to each component, the table also serves as a basis for comparing the relative cost  of the proposed architectures.

\section{Power Allocation and Hybrid beamforming}

The optimal design of power allocation and hybrid combiners in multi-user MIMO systems aims at minimizing inter-user interference and maximizing spectral efficiency (SE). This problem becomes increasingly complex in multi-band scenarios where multiple sub-bands are used concurrently for different users. 
Let $\{\lambda_k\}_{k \in \mathcal{K}}$ denote the power allocations for each user equipment (UE) $k$. To maximize the total sum rate, we jointly optimize the power allocation $\{\lambda_k\}$ and hybrid beamforming vectors $\{\mathbf{w}_k\}$ for each UE.
The unconstrained spectral efficiency for the $k$-th UE in the $c$th sub-band $B_c$ can be expressed as:
\begin{equation}
    \eta_k[c] = \mu \, \mathbb{E}\left[\log_2 \left(1 + \text{SINR}_k[c] \right)\right],
\end{equation}
where $\mu \leq 1$ accounts for communication overhead, such as pilot training for channel estimation, while $\text{SINR}_k$ is computed as:
\begin{equation}
    \text{SINR}_k[c] = \frac{\left\|\mathbf{w}_k^\mathrm{H}[c] \, \mathbf{h}_k[c] \, s_k\right\|^2}{\left\|\sum_{\substack{\ell  \in \mathcal{K}_c \\ \ell \neq k}} \mathbf{w}_k^\mathrm{H} [c]\, \mathbf{h}_\ell [c]\, s_\ell \right\|^2+ \left\|\mathbf{w}_k^\mathrm{H}[c] \, \mathbf{n}[c]\right\|^2},
\end{equation}
where interference affecting UE $k$ is due only to the users allocated on the same sub-band $\ell \in \mathcal{K}_c$.
The proposed objective consists of maximizing the total sum rate:
\begin{equation}
    \eta_\mathrm{Tot} = \sum_c \sum_{k \in \mathcal{K}_c} \eta_k[c].
\end{equation}
The optimization problem can be formulated as:
\begin{subequations}
\begin{align}\label{eq:optGeneral}
    \max_{\{\mathbf{w}_k, \lambda_k\}_{k \in \mathcal{K}}} \quad & \sum_c \sum_{k \in \mathcal{K}_c} \eta_k[c]. \\
    \text{s.t.} \quad & 0 \leq \lambda_k \leq P_{\text{max}}, \quad \forall k \in \mathcal{K}, \label{subeq:powerConstraint} \\
    & \mathbf{W}_{\text{RF}} \in  \mathcal{W}_\mathrm{RF}. \label{subeq:RFConstraint}
\end{align}
\end{subequations}
where $\mathcal{W}_\mathrm{RF}$ is a set of feasible RF combiners.
The constraint in \eqref{subeq:powerConstraint} limits the power allocated to each UE. The constraint in \eqref{subeq:RFConstraint} captures the inherent hardware limitations of the RF combiner, imposing that all its elements maintain a uniform norm, as dictated by the analog phase shifter implementation \cite{9475519,6717211}. This constraint makes the problem non-convex. However, the problem remains tractable using suitable numerical methods, with sub-optimal solutions. We employ an alternating maximization (AM) strategy, which has been demonstrated to converge to the global optimum for this class of problems \cite{bezdek2002some}. In each iteration of the algorithm, we first hold the power constant and focus on optimizing the beamforming vectors (as detailed in Section V.A). Following this, we fix the beamforming vectors and proceed with optimizing the power allocation (as described in Section V.B). 

\subsection{Optimizing the beamforming Matrix}

We now address the optimization of the hybrid combiners under fixed power allocation. The problem can be expressed as:
\begin{subequations}
\begin{align}\label{eq:optHybridCombiner}
    \max_{\{\mathbf{w}_k\}_{k \in \mathcal{K}}} \quad & \sum_c \sum_{k \in \mathcal{K}_c} \eta_k[c] \\
    \text{s.t.} \quad & \mathbf{W}_{\text{RF}} \in  \mathcal{W}_\mathrm{RF}. 
\end{align}
\end{subequations}
This problem is non-convex due to the norm constraint on $\mathbf{W}_{\text{RF}}$ elements and the non-convex objective function involving the SINR. 
To simplify this, we use a two-step approach similar to \cite{6717211}. First, we relax the hybrid constraint and optimize the combiners as fully digital, treating each $\mathbf{w}_k$ as unconstrained. Second, we reconstruct the hybrid combiner using the orthogonal matching pursuit (OMP) algorithm.

\subsubsection{Optimal Unconstrained Combiner}

This can be reformulated as a minimization of the mean squared error (MSE) \cite{guo2005mutual}, which is equivalent to maximizing the SINR, or equivalently as
\begin{equation}
    \min_{\mathbf{w}_k} \quad \frac{1}{\text{SINR}_k[c]}.
\end{equation}
The optimal solution is given by the Linear Minimum Mean Squared Error (LMMSE) combiner:
\begin{equation}\label{eq:optCombiner}
    \mathbf{w}_k^{\text{LMMSE}} = \left( \sigma_n^2 \mathbf{I}_{N_A} + \sum_{\substack{\ell  \in \mathcal{K}_c \\ \ell \neq k}}  \lambda_\ell \widehat{\mathbf{h}}_\ell \widehat{\mathbf{h}}_\ell^\mathrm{H} \right)^{-1} \widehat{\mathbf{h}}_k,
\end{equation}
where $\widehat{\mathbf{h}}_\ell$ represents the estimated channel vector for the $\ell$th UE. Instead of \eqref{eq:optCombiner}, one can use suboptimal solutions methods such as maximum ratio combining (MRC) and zero force combining (ZFC).

\subsubsection{Hybrid Combiner Construction}

Once the optimal full digital combiner $\mathbf{W}_{\text{opt}}$ is determined (either LMMSE, ZFC, or MRC), the hybrid combiner is derived using the OMP algorithm. Let $\mathbf{D} \in \mathbb{C}^{N_a \times L}$ be a predefined beamforming codebook, e.g., a discrete Fourier transform (DFT) codebook \cite{9475519}. The analog and digital combiners, $\mathbf{W}_{\text{RF}}$ and $\mathbf{W}_{\text{BB}}$, are computed through OMP, described in Algorithm \ref{alg:OMP}.

\begin{algorithm}[b!]
\footnotesize
\caption{Orthogonal Matching Pursuit (OMP) Algorithm}
\begin{algorithmic}[1]\label{alg:OMP}
\STATE \textbf{Input:} $\mathbf{W}_{\text{opt}}, \mathbf{D}$
\STATE \textbf{Output:} $\mathbf{W}_{\text{RF}}, \mathbf{W}_{\text{BB}}$
\STATE Initialize: Residual $\mathbf{W}_\text{res} = \mathbf{W}_{\text{opt}}$
\FOR{$i = 1$ to $N_\text{RF}$}
    \STATE $\mathbf{\Gamma} = \mathbf{W}_\text{res}^\mathrm{H} \mathbf{D}$
    \STATE $q_\text{max} = \arg \max_q \left( \mathbf{\Gamma}^\mathrm{H} \mathbf{\Gamma} \right)_{q,q}$
    \STATE $\mathbf{W}_{\text{RF}} = \left[\mathbf{W}_{\text{RF}} | \mathbf{D}^{(q_\text{max})}\right]$
    \STATE $\mathbf{W}_{\text{BB}} = \left( \mathbf{W}_{\text{RF}}^\mathrm{H} \mathbf{W}_{\text{RF}} \right)^{-1} \mathbf{W}_{\text{RF}}^\mathrm{H} \mathbf{W}_\text{opt}$
    \STATE Update residual: $\mathbf{W}_\text{res} = \frac{\mathbf{W}_\text{opt} - \mathbf{W}_{\text{RF}} \mathbf{W}_{\text{BB}}}{\left\|\mathbf{W}_\text{opt} - \mathbf{W}_{\text{RF}} \mathbf{W}_{\text{BB}}\right\|_\text{F}}$
\ENDFOR
\end{algorithmic}
\end{algorithm}

This two-step approach ensures that the hybrid combiner closely approximates the performance of the optimal full-digital combiner while adhering to the practical hardware constraints \cite{6717211}.

\subsection{Optimizing Power Allocation}

Given fixed hybrid beamforing vectors \(\{\mathbf{w}_k\}\) obtained by Algorithm \ref{alg:OMP}, the power optimization problem can be expressed as:
\begin{equation}\label{eq:subOpt1}
    \max_{\{\lambda_k\}} \sum_c \sum_{k \in \mathcal{K}_c} \eta_k[c] \quad \text{s.t.} \quad 0 \leq \lambda_k \leq P_{\text{max}}, \quad \forall k \in \mathcal{K}.
\end{equation}
The SINR expression in the objective function is generally a fractional quadratic function of the power variables \(\{\lambda_k\}\), which makes the overall problem non-convex.
However, a common simplification is to treat interference from other users as fixed noise \cite{4357612}. Under this approximation, the SINR for each user depends only on power allocation set \(\{\lambda_k\}\), thus transforming the problem into a tractable convex optimization problem.

We adopt the method of Lagrange multipliers to handle the per-user power constraints. The Lagrangian is
\begin{equation}
    \mathcal{L}(\{\lambda_k\}, \{\nu_k\}, \{\gamma_k\}) = \sum_{k \in \mathcal{K}} \eta_k - \sum_{k \in \mathcal{K}} \nu_k (\lambda_k - P_{\text{max}}) + \sum_{k \in \mathcal{K}} \gamma_k \lambda_k,
\end{equation}
where \(\nu_k \geq 0\) and \(\gamma_k \geq 0\) are  associated to the constraints \(\lambda_k \leq P_{\text{max}}\) and \(\lambda_k \geq 0\), respectively.
The optimal power allocations \(\{\lambda_k\}\) are determined by solving the Karush-Kuhn-Tucker (KKT) stationarity condition \cite{4357612}:
\begin{equation}
    \frac{\partial \eta_k}{\partial \lambda_k} - \nu_k + \gamma_k = 0, \quad \forall k \in \mathcal{K}.
\end{equation}
given the primal and dual visibility conditions $0 \leq \lambda_k \leq P_{\text{max}}$, $\nu_k \geq 0$ and $\gamma_k \geq 0$, $\forall k \in \mathcal{K}$, respectively.

This simplified problem can be solved using the water-filling algorithm. The optimal power allocation for each user \(k\) is given by:
\begin{equation}\label{eq:waterfilling}
    \lambda_k = \left[ \frac{\mu}{\ln(2) \, \nu_k} - \frac{\sigma_n^2 + \sum_{\substack{\ell \in \mathcal{K}_c \\ \ell \neq k}} \lambda_\ell \left|\mathbf{w}_k^\mathrm{H} \mathbf{h}_k^{(c)}\right|^2}{\left|\mathbf{w}_k^\mathrm{H} \mathbf{h}_k^{(c)}\right|^2} \right]^+,
\end{equation}
where \([\cdot]^+\) denotes the projection onto non-negative values, ensuring \(\lambda_k \geq 0\), and $\gamma_k \geq 0$. The proposed algorithm operates iteratively: each user updates their transmit power \(\lambda_k\) using the water-filling expression \eqref{eq:waterfilling}, while the interference is recalculated at each step. This process continues until convergence or a maximum number of iterations is reached.

\subsection{Alternating Maximization Algorithm}

The complete AM algorithm alternates between optimizing the hybrid combiners and the power allocation until convergence. The steps of the AM algorithm are in Algorithm \ref{alg:AM},
\begin{algorithm}[t!]
\footnotesize
\caption{Alternating Maximization (AM) Algorithm for Power Allocation and Hybrid beamforing}\label{alg:AM}
\begin{algorithmic}[1]
\STATE \textbf{Input:} $\{\widehat{\mathbf{h}}_k\}$, $\sigma_n^2$, $P_{\text{max}}$
\STATE \textbf{Output:} $\{\lambda_k\}$, $\{\mathbf{W}\}$
\STATE Initialize: $\lambda_k^{(0)} \leftarrow P_{\text{max}}$
\FOR{$n = 1$ to $L_{\text{max}}$}
    \STATE \textbf{Step 1:} Compute optimal combiner using \eqref{eq:optCombiner},
    \STATE \textbf{Step 2:} Compute hybrid combiner using Algorithm \ref{alg:OMP},
    \STATE \textbf{Step 3}: Compute optimal power using \eqref{eq:waterfilling}
    
    \IF{$\|\eta_\mathrm{Tot}^{(n)} - \eta_\mathrm{Tot}^{(n-1)}\|_2 < \epsilon$}
        \STATE \textbf{Break}
    \ENDIF
\ENDFOR
\end{algorithmic}
\end{algorithm}
where $\epsilon$ defines the stop condition for the AM algorithm.

\begin{table}[t]
    \centering
    \footnotesize
    \renewcommand{\arraystretch}{1.4} 
    \setlength{\tabcolsep}{10pt} 
    \begin{tabular}{|>{\bfseries}l|c|l|}
        \hline
        \rowcolor{gray!20} 
        \textbf{Parameter}        & \textbf{Symbol}         & \textbf{Value} \\ 
        \hline
        Channel model             & -                       & 3GPP UMI [REF] \\
        Number of antennas        & $N_a$                   & 16 (for each sub-band)\\
        Number of RF chains       & $N_{\text{RF}}$         & 8 (for each sub-band) \\
        Maximum transmit power    & $P_{\text{max}}$        & 10 dBm \\
        Noise power               & $\sigma_n^2$            & $-90$ dBm \\
        Block Length              & $\tau_b$                & 200 \\
        \hline
    \end{tabular}
    \caption{Simulation parameters.}
    \label{tab:simulation_parameters}
\end{table}

\section{Simulation Results}

In this section, we analyze how different architectures, beamforming methods, and power allocation strategies affect the system sum rate. We examine MRC, R-ZFC, and LMMSE and distinguish between FD and HAD architectures with SRF or DRF designs.

To account for channel estimation, we model the estimated channel as $\widehat{\mathbf{h}}_\ell = \mathbf{h}_\ell + \mathbf{n}_h$, where $\mathbf{n}_h \sim \mathcal{N}(\mathbf{0}, \sigma_{n_h}^2 \mathbf{I}_{N_A})$ denotes the channel estimation error when ideal pilots are employed. This model complies with the Least Squares channel estimation method \cite{morelli2001comparison}, where $\sigma_{n_h}^2$ corresponds to the mean square error of the estimated channel. The cost for channel estimation is included in the spectral efficiency parameter $\mu = 1 - \tau_p/\tau_b$, with $\tau_b$ being the block fading length, and $\tau_p$ denoting the pilot length, which is proportional to the number of UEs served simultaneously in a sub-band.
Unless stated otherwise, the simulations utilize the parameters listed in Table~\ref{tab:simulation_parameters}. Key hardware parameters are presented in Table~\ref{tab:hardware_parameters}, which can be found in the datasheets of commercially available components.

\begin{table}[t]
    \centering
    \footnotesize
    \renewcommand{\arraystretch}{1.4} 
    \setlength{\tabcolsep}{8pt} 
    \caption{Hardware parameters with noise figures, gains, static power, and dynamic power coefficients.}
    \begin{tabular}{|>{\bfseries}l|c|c|c|c|}
        \hline
        \rowcolor{gray!20} 
        \textbf{Component}         & \textbf{NF} & \textbf{G} & \(\mathbf{P^s}\) \textbf{(dBW)} & \(\mathbf{\kappa}\) \textbf{(dBW/Hz)} \\ 
        \hline
        LNA  & 1  & 18  & -20  & -123  \\
        RF Filter                 & 2   & -0.2 & -   & -  \\
        IF Filter                 & 1   & -0.2 & -   & -  \\
        RF-IF Mix            & 22  & -0.5 & -23  & -123  \\
        IF-BB Mix            & 18  & -0.5 & -   & -  \\
        ADC                       & 22  & 0    & -23  & -114  \\
        D/C                   & 3   & -0.5 & -   & -  \\
        PS             & 3   & 0.2  & -   & -  \\
        \hline
    \end{tabular}
    \label{tab:hardware_parameters}
    \vspace{-.5cm}
\end{table}

The results show that LMMSE consistently outperforms R-ZFC and MRC in all scenarios. FD architectures outperform HAD solutions due to their higher degrees of freedom. At the same time, FI classes achieve higher rates than FP classes, more than $4\times$ in the considered setup. 

\begin{figure}[b!]
\vspace{-0.5cm}
    \centering
    \includegraphics[width=0.95\linewidth]{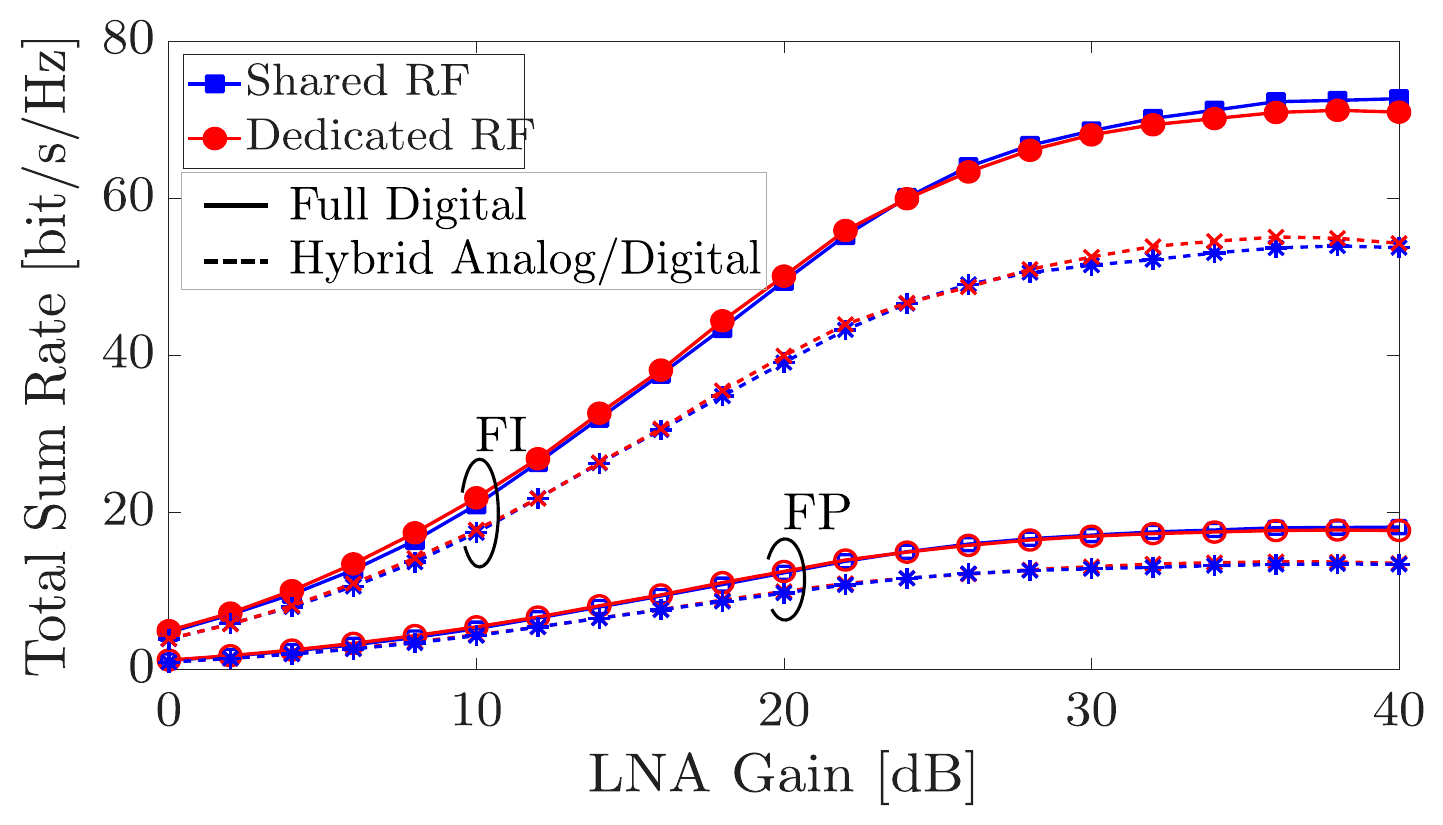}
    \caption{Impact of LNA gain of the total sum rate considering all analyzed architectures. The result assume perfect CSI, $P_\mathrm{max} = 10$ dBm for UE, and $\sum_c K_c = 16$ UEs.}
    \label{fig:RvsLNA}
\end{figure}

The results presented in Fig.~\ref{fig:RvsLNA} show the total sum rate for both FD and HAD architectures as a function of the LNA gain. As the LNA gain increases, the noise figure introduced by later stages is reduced, leading to a higher SNR. 
Generally, increasing the LNA gain results in an improvement in the sum rate, up until a saturation point is reached. Beyond this point, the noise figure is primarily determined by the LNA itself. For the considered setup, the saturation point is observed at approximately 35 dB for FD and 25 dB for HAD.
Above the saturation point, the system transitions from being noise-limited to interference-limited.

\begin{figure}[b!]
\vspace{-0.5cm}
    \centering
    \subfloat[Total sum rate for FD architecture]{\includegraphics[width=0.95\linewidth]{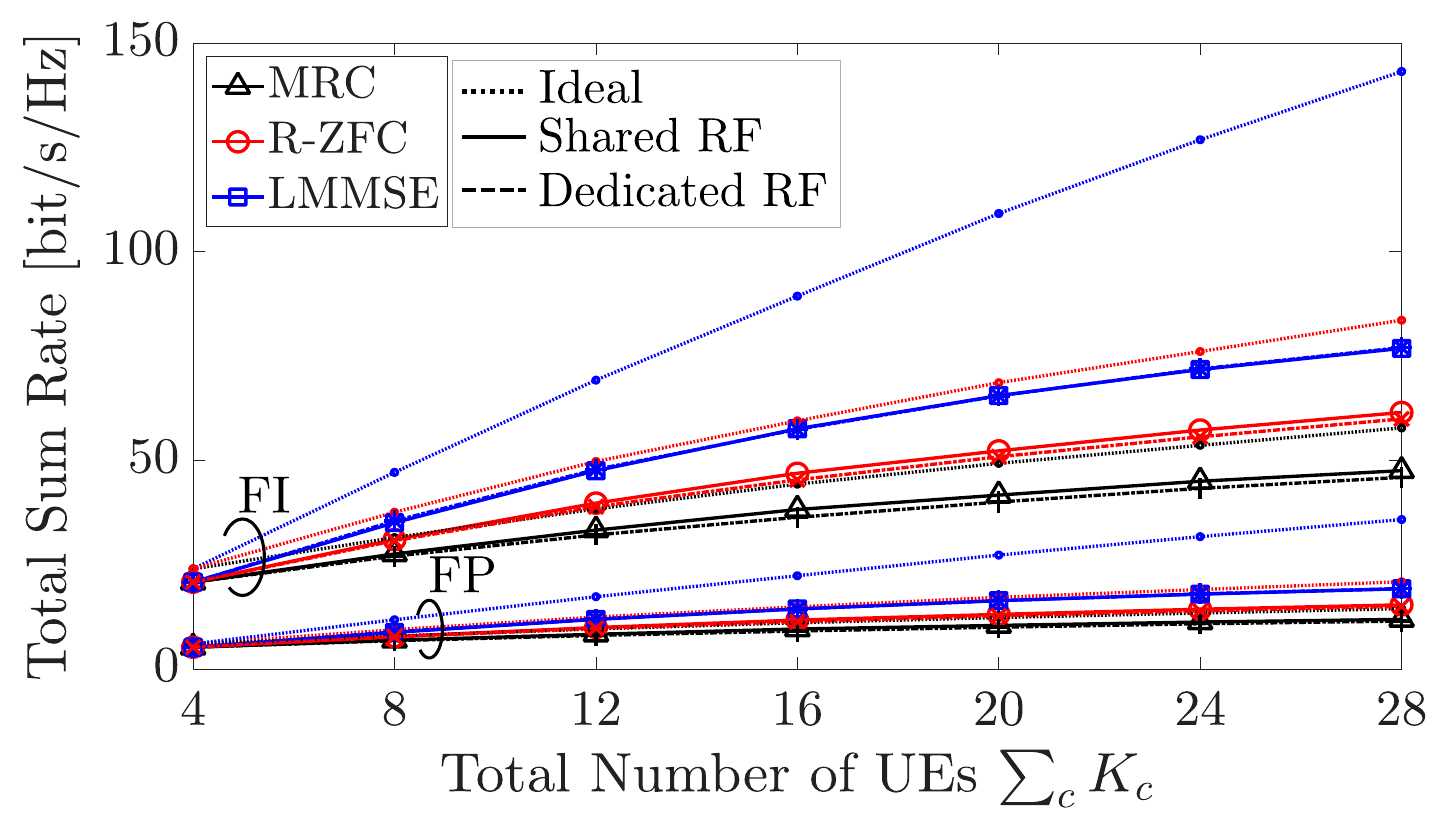}\label{subfig:UEFDS}} \\
    \subfloat[Total sum rate for HAD architecture]{\includegraphics[width=0.95\linewidth]{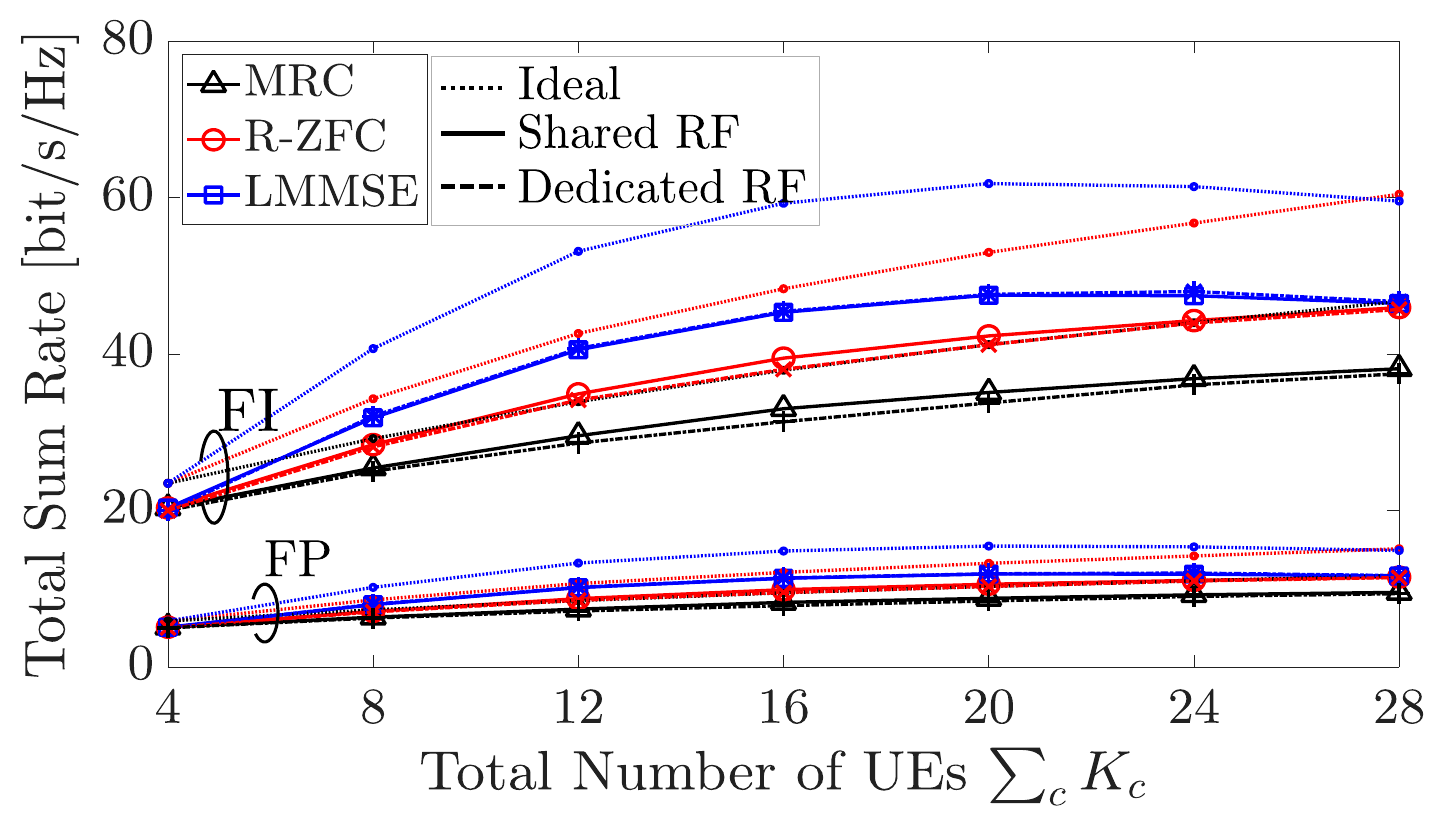}\label{subfig:UEHADS}} 
    \caption{Total sum rate vs number of UEs for (a) FD and (b) HAD architectures. No hardware impairments (dash-dot lines), hardware impairments with SRF design (solid lines), and hardware impairments with DRF design (dashed lines). The results are based on optimal power allocation, with $P_\mathrm{max} = 10$ dBm, and perfect CSI.}
    \label{fig:RvsUE}
\end{figure}

The FD architecture is more sensitive to hardware impairments than the HAD architecture, and its performance degrades more noticeably when low-gain LNAs are used. When considering low-gain LNAs, DRF architecture performs better because of the reduced number of components in the receiver chain. However, as the LNA gain increases, the SRF architecture shows better performance. This trend does not hold when the number of users is low, as can be observed also in Fig. \ref{fig:RvsUE}, and it can be attributed to the greater ability of the SRF architecture to mitigate interference. The improved interference management is made possible by the antenna structure in the SRF case, where antennas are positioned at half the wavelength (\(\lambda/2\)) relative to the lowest carrier frequency (6 GHz). At higher frequencies, the array spacing becomes greater than \(\lambda_c/2\), resulting in improved selectivity compared to the DRF solution, which uses different antennas for each carrier with each antenna adapted to the individual carrier's.

\begin{figure}[t!]
    \centering
    \includegraphics[width=0.95\linewidth]{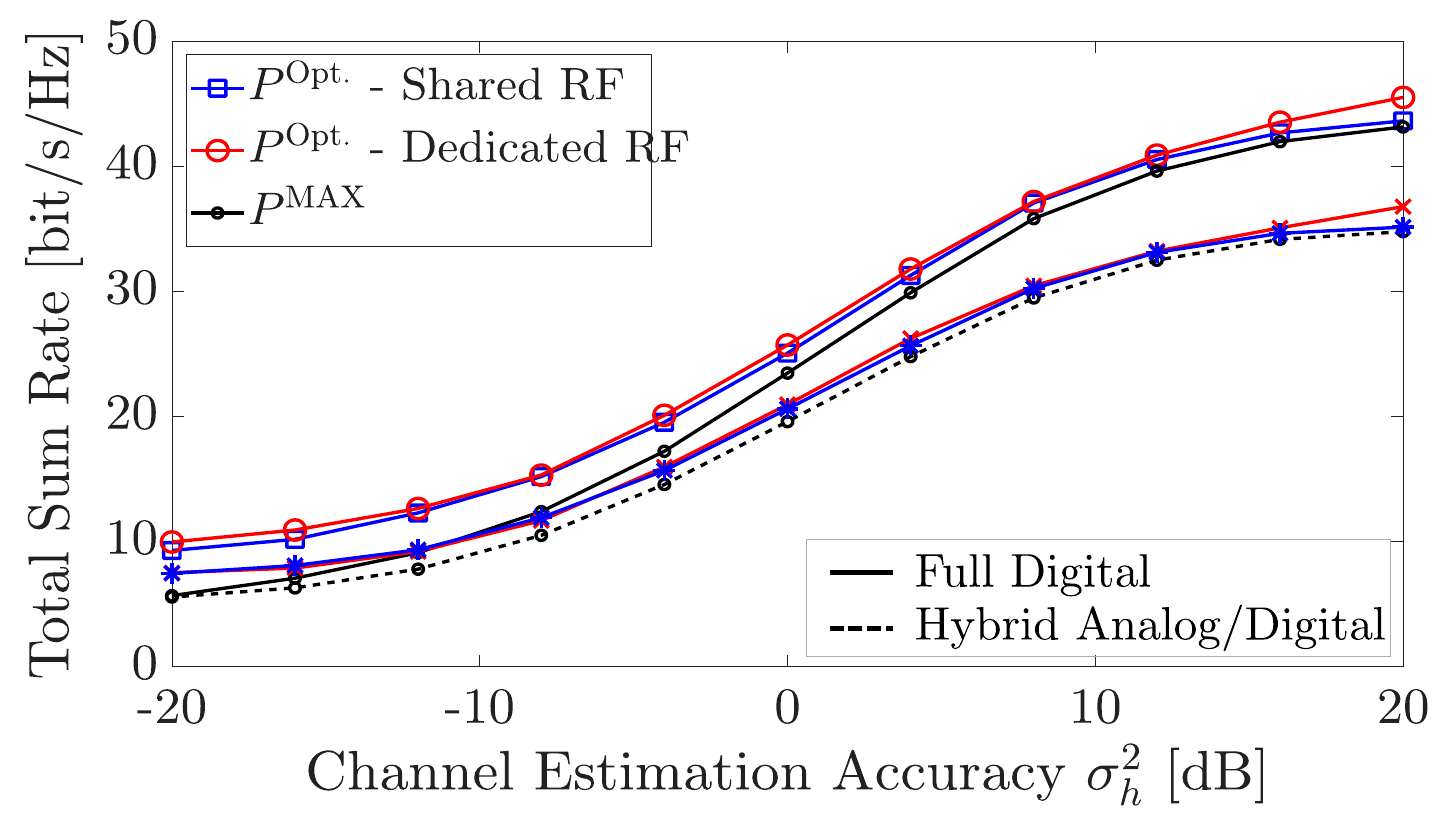}
    \caption{Impact of channel estimation accuracy on the total sum rate, comparing optimal power allocation and max power. This result consider $P_\mathrm{max} = 10$ dBm, and $\sum_c K_c = 16$ UEs.}
    \label{fig:RvsH}
    \vspace{-.5cm}
\end{figure}

Figure \ref{fig:RvsH} depicts the impact of imperfect CSI by varying the MSE of channel estimation. The analysis includes both FD and HAD architectures, considering both SRF and DRF design, as well as a benchmark scenario with maximum power without optimization.
As expected, imperfect CSI leads to performance degradation in all configurations, regardless of the architecture considered. However, as channel estimation becomes less accurate, power optimization strategies become increasingly important. Both FD and HAD architectures, despite their different structural approaches, show comparable performance in such conditions, emphasizing the role of power control in overcoming CSI imperfections. 

The results in Fig.~\ref{fig:RvsUE} show the total sum rate as the number of UEs increases. Figures~\ref{subfig:UEFDS} and \ref{subfig:UEHADS} depict the results for FD and HAD architectures, respectively. The analysis compares three scenarios: no hardware impairments (dash-dot lines), hardware impairments in SRF design (solid lines), and hardware impairments with DRF design (dashed lines), assuming optimal power allocation in all cases.
As the number of UEs grows, the sum rate initially increases for all configurations and beamforming techniques, as the system effectively manages interference. However, the system eventually reaches a saturation point where adding more UEs no longer improves performance. Beyond this point, the sum rate decreases because the system becomes interference-limited. This behavior is evident in both FD and HAD architectures, with HAD systems reaching the saturation point earlier due to their limited number of RF chains. 

\begin{figure}[t!]
    \centering
    \subfloat[Total sum rate for Full Digital (FD) architecture]{\includegraphics[width=0.95\linewidth]{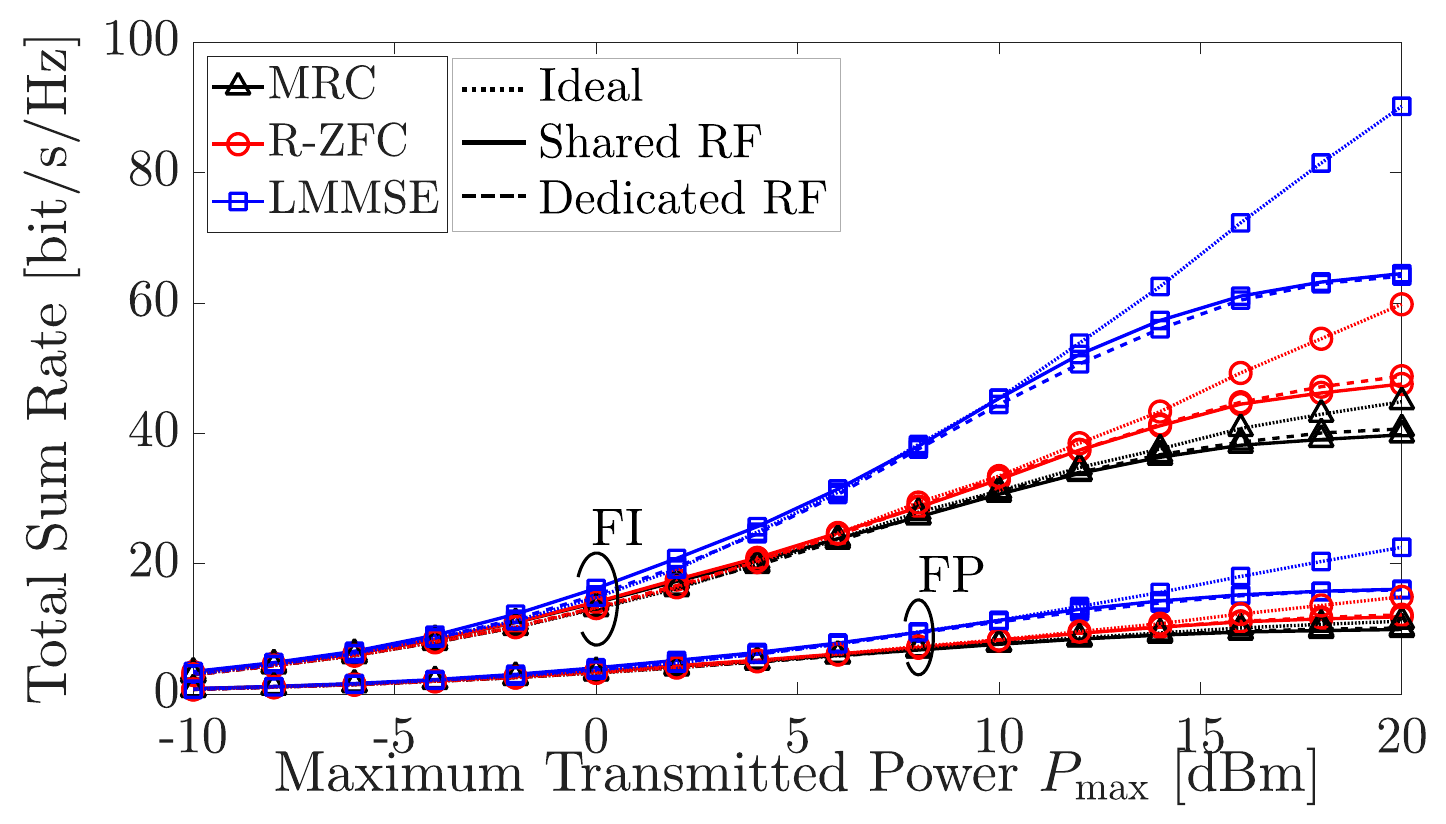}\label{subfig:PtFDS}} \\
    \subfloat[Total sum rate for Hybrid Analog/Digital (HAD) architecture]{\includegraphics[width=0.95\linewidth]{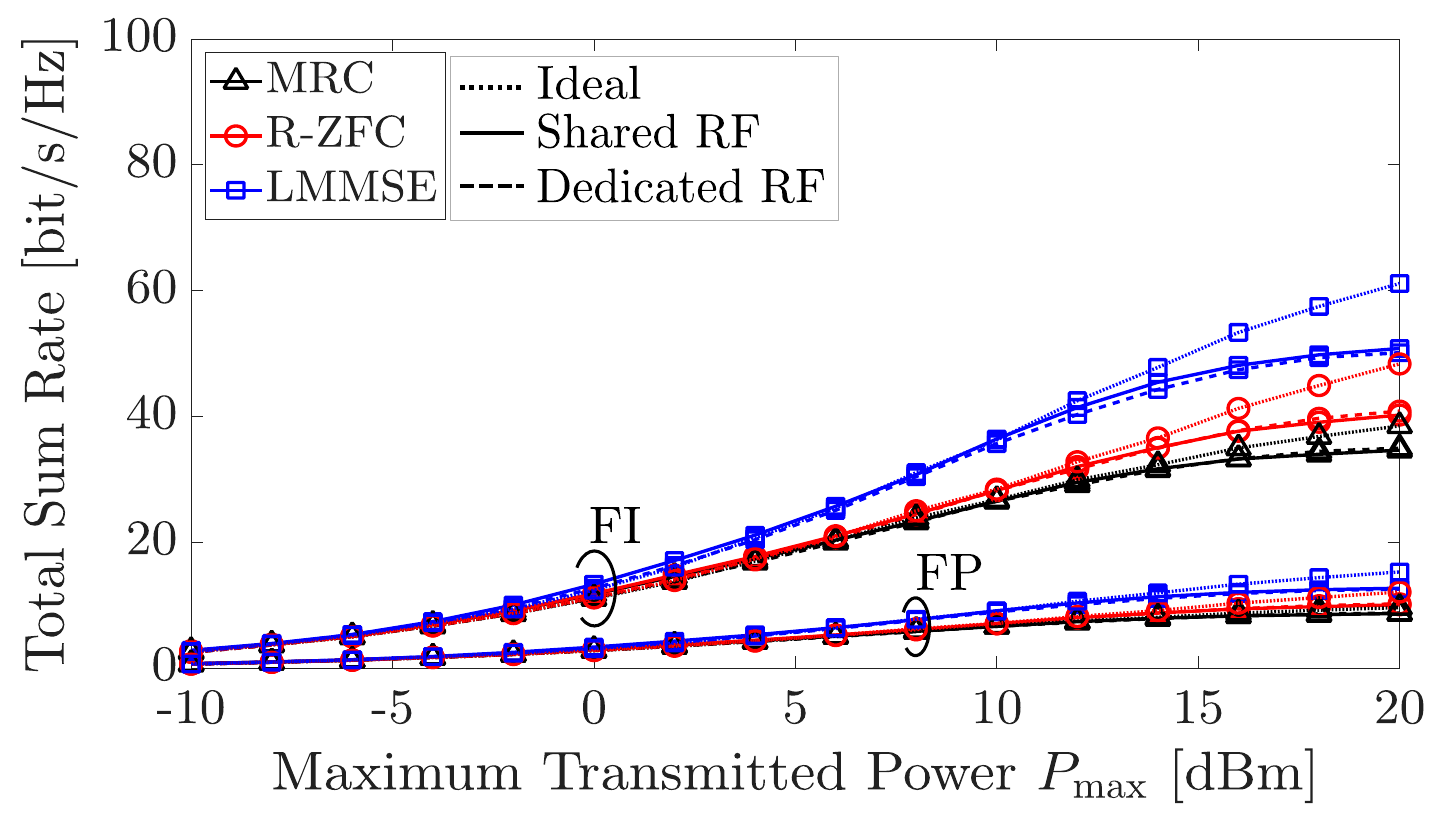}\label{subfig:PtHADS}} \\
    \caption{Impact of increasing the maximum transmission power of UEs on the total sum rate for (a) FD (a) and (b) HAD architectures. No hardware impairments (dash-dot lines), hardware impairments with SRF (solid lines), and hardware impairments with DRF (dashed lines). The results are based on optimal power allocation, perfect CSI, and $\sum_c K_c = 16$ UEs.}
    \label{fig:RvsPt}
    \vspace{-.5cm}
\end{figure} 

Hardware impairments affect the overall performance in all scenarios, reducing the achievable sum rate. Among the beamforming methods, LMMSE is the most sensitive to hardware impairments, suffering the largest degradation in performance, while MRC is less affected due to its simpler processing, albeit achieving lower performance overall. 

Figure \ref{fig:RvsUE} shows the effect of increasing the maximum transmission power of UEs on the total sum rate of the system. Figures~\ref{subfig:PtFDS} and \ref{subfig:PtHADS} depict the results for FD and HAD architectures, respectively. Three different use cases are considered: the first with no hardware impairments (dash-dot lines), the second with hardware impairments using SRF (solid lines), and the third with hardware impairments using DRF (dashed lines).
As the available power at the UE increases, the sum rate improves for all architectures and beamforming methods. In the absence of hardware impairments, both FD and HAD architectures show improvements, with FD benefiting more due to its higher degrees of freedom for interference management. The HAD architecture shows a performance increase up to a certain point, after which it saturates due to limited interference mitigation capabilities. When hardware impairments are introduced, both FD and HAD architectures reach saturation earlier than in the case without impairments. In these scenarios, the system becomes noise-limited, and DRF solutions perform slightly better than SRF ones, especially for HAD architectures.

\begin{figure}[t]
    \centering
    \subfloat[Frequency Integrated (FI) architectures]{
        \includegraphics[width=0.45\textwidth]{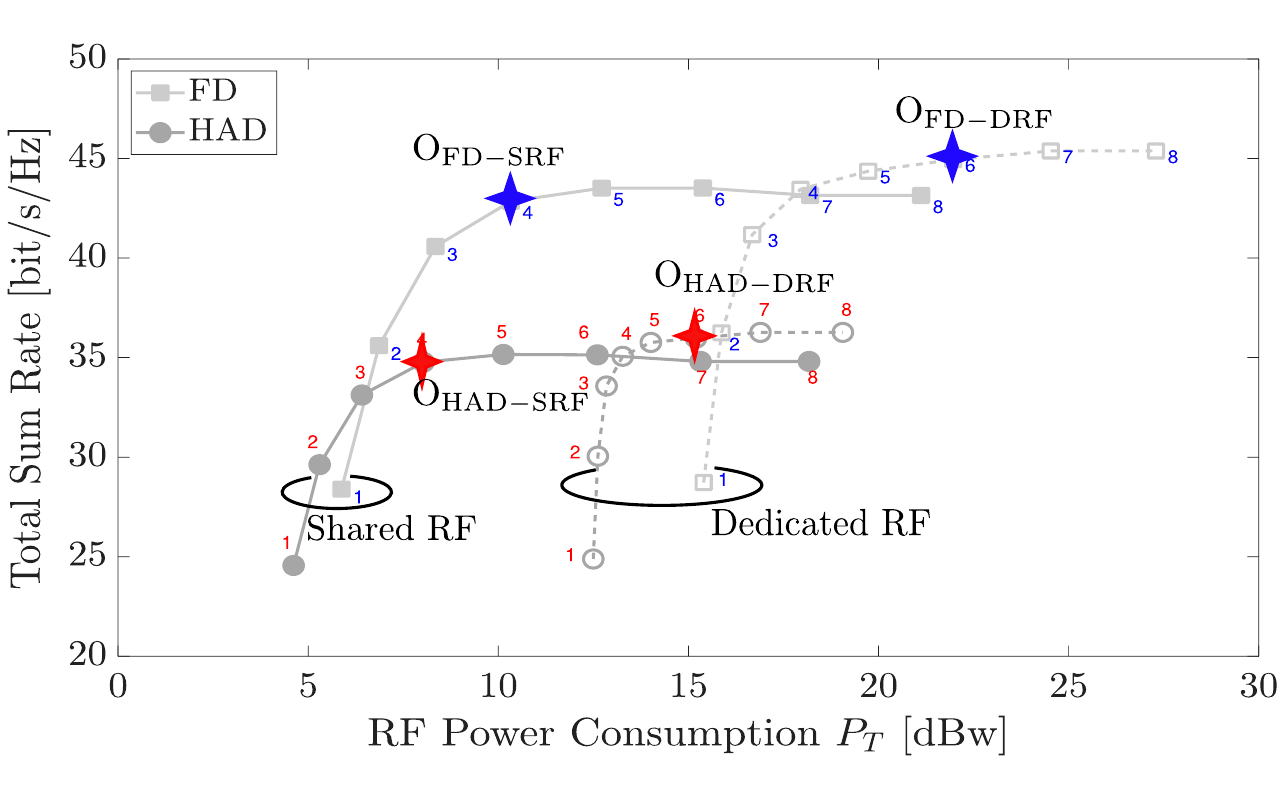}
        \label{fig:FI_power_vs_SE}
    }
    \\
    \subfloat[Frequency Partitioned (FP) architectures]{
        \includegraphics[width=0.45\textwidth]{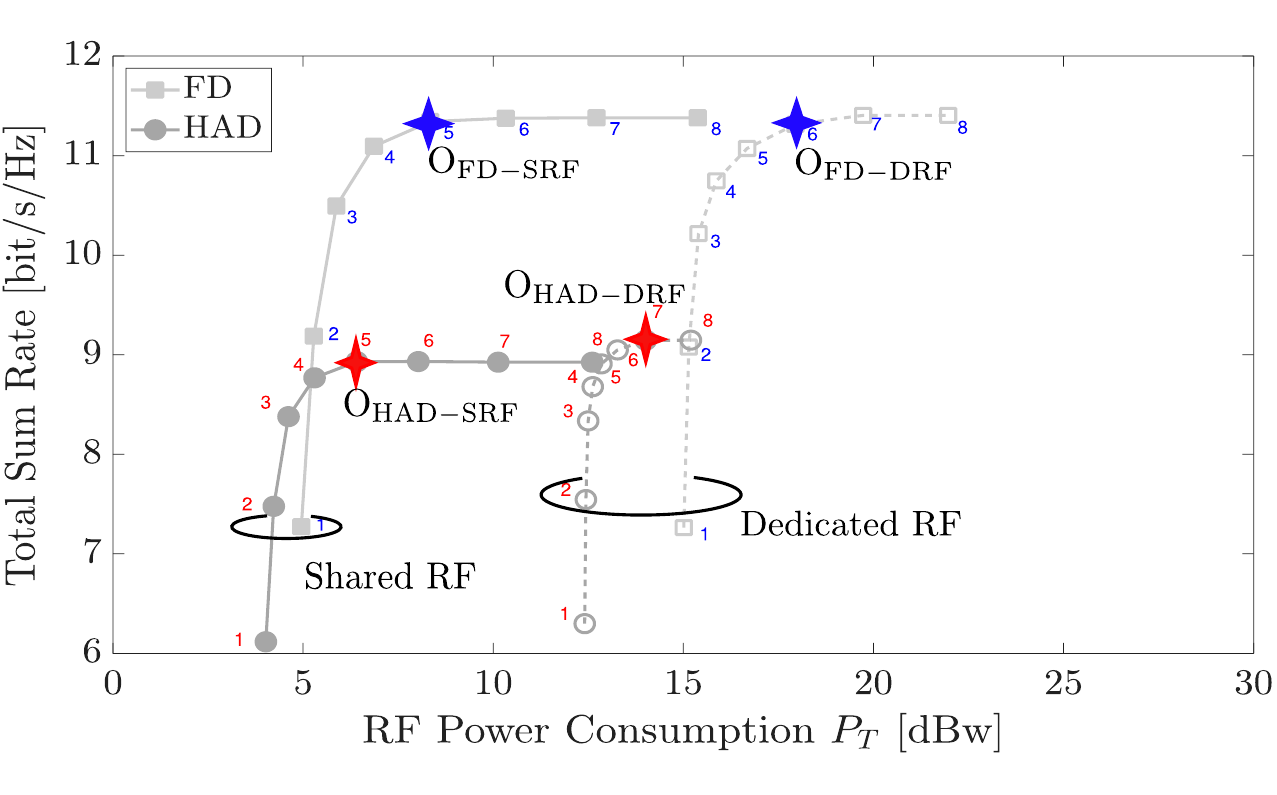}
        \label{fig:FP_power_vs_SE}
    }
    \caption{Spectral efficiency vs. power consumption for (a) FI and (b) FP architectures with varying ADC resolution. Full-digital are with square markers and hybrid analog-digital are with circular markers. The numbers next to markers represent the ADC resolution in bits, and the optimal sum-rate/RF power consumption trade-off are highlighted by star marker.}
    \label{fig:power_vs_SE}
    \vspace{-.5cm}
\end{figure}

\begin{table}[b!]
\centering
\caption{Optimal ADC bit configuration and corresponding spectral efficiency (SE) and power consumption (\(P_T\)).}
\renewcommand{\arraystretch}{1.5} 
\setlength{\tabcolsep}{6pt} 
\begin{tabular}{|c|c|c|c|c|}
\hline
\rowcolor{gray!20} 
\textbf{Class} & \textbf{Architecture} & \textbf{$N_\mathrm{bits}^\mathrm{Opt}$} & SE$^\mathrm{Opt}$ \textbf{(bps/Hz)} & \(P_T^\mathrm{Opt}\) \textbf{(dBW)} \\ 
\hline \hline
\multirow{4}{*}{\textbf{FI}}  
& FD-SRF  & 4  & 42.8  & 10.32  \\ \cline{2-5} 
& HAD-SRF & 4  & 34.7  & 8.0    \\ \cline{2-5}  
& FD-DRF  & 6  & 44.9  & 21.9   \\ \cline{2-5}  
& HAD-DRF & 6  & 35.9  & 15.1   \\  
\hline \hline
\multirow{4}{*}{\textbf{FP}}  
& FD-SRF  & 5  & 11.3  & 8.3    \\ \cline{2-5}  
& HAD-SRF & 5  & 8.9   & 6.4    \\ \cline{2-5}  
& FD-DRF  & 6  & 11.3  & 18.9   \\ \cline{2-5}  
& HAD-DRF & 7  & 9.1   & 14.0   \\  
\hline
\end{tabular}
\label{tab:optimalADC}
\end{table}

To evaluate the trade-off between spectral efficiency and power consumption, we analyze the total sum rate as a function of the total consumed power for different ADC resolutions. The experiment considers both FI and FP architectures, as shown in Figs.~\ref{fig:FI_power_vs_SE} and~\ref{fig:FP_power_vs_SE}, respectively. Each figure presents four curves: continuous lines represent SRF architectures, while dashed lines correspond to DRF. Additionally, square markers indicate FD architectures, whereas circular markers refer to HAD designs. The numbers next to each marker denote the ADC resolution in bits.

The results reveal consistent trends across all architectures. Initially, increasing the number of ADC bits significantly enhances spectral efficiency without a substantial increase in power consumption. However, beyond a certain saturation point, denoted as the optimal 
operating point, also shown in Table \ref{tab:optimalADC}, further increasing the ADC resolution results in a significant rise in power consumption without any notable gain in spectral efficiency. Furthermore, HAD architectures consistently consume less power than FD architectures but at the cost of reduced spectral efficiency. In comparison, DRF architectures exhibit significantly higher power consumption than their shared counterparts, despite delivering comparable spectral efficiency. This discrepancy is attributed to the increased static power consumption of dedicated RF components. 
Lastly, FP architectures exhibit slightly better power efficiency than FI architectures, as observed in Figure~\ref{fig:power_vs_SE}; however, they achieve lower spectral efficiency due to the sequential access of sub-bands.

\section{Conclusions}

This paper investigates architectural and signal processing strategies for multi-band and multi-user systems for upper mid-band (FR3) communication systems, emphasizing the trade-off between spectral efficiency and power consumption as a key design challenge. By examining two primary architectural classes—\textit{Frequency Integrated (FI)} and \textit{Frequency Partitioned (FP)}—this study provides critical insights into the optimal design of multi-band and multi-user FR3 systems.

The key findings of this work are summarized as follows:

\begin{itemize}
    \item \textbf{FI vs. FP Architectures:} FI architectures, which process multiple sub-bands simultaneously, achieve significantly higher spectral efficiency, exceeding $4\times$ the performance of FP designs, which dynamically schedule sub-band access. However, this comes at the cost of slightly higher power consumption.

    \item \textbf{SRF vs. DRF Designs:} Shared RF chain (SRF) and dedicated RF chain (DRF) architectures achieve comparable spectral efficiency. However, SRF designs consume approximately half the power of DRF configurations, making them a more energy-efficient choice in the considered setup.

    \item \textbf{HAD vs. FD Architectures:} Fully digital (FD) architectures provide higher spectral efficiency than hybrid analog-digital (HAD) systems but at the cost of significantly increased power consumption. HAD architectures offer a more favorable balance between performance and energy efficiency, particularly in SRF configurations. The spectral efficiency gap narrows as the number of UEs served decreases.

    \item \textbf{Impact of beamforming and Power Allocation:} The study evaluates dedicated beamforming techniques, including LMMSE, R-ZFC, and MRC, alongside optimal power allocation strategies, demonstrating their role in improving system efficiency.

\end{itemize}

These findings provide valuable engineering guidelines for next-generation FR3 communication systems, assisting in optimizing the balance between spectral efficiency and power consumption in multi-band and multi-user scenarios. Nevertheless, several challenges remain. In particular, future research should tackle the complexities of \textit{near-field communication} and investigate methods for \textit{mixed near-field and far-field} processing techniques. 
\bibliographystyle{IEEEtran}
\bibliography{bibliography}

\end{document}